\pdfoutput=1

\documentclass[11pt,twoside,a4paper,cmspaper,final,collab]{cms-tdr}
\def\svnVersion{307607}\def\cmsCernNo{2013-037}\def\cmsCernDate{\today}\def\cmsMessage{Published in the Journal of High Energy Physics as \href{http://dx.doi.org/10.1007/JHEP10(2015)128}{\doi{10.1007/JHEP10(2015)128.}}}

\begin{document}\cmsNoteHeader{SMP-14-005}

\hyphenation{had-ron-i-za-tion}
\hyphenation{cal-or-i-me-ter}
\hyphenation{de-vices}
\RCS$Revision: 304054 $
\RCS$HeadURL: svn+ssh://svn.cern.ch/reps/tdr2/papers/SMP-14-005/trunk/SMP-14-005.tex $
\RCS$Id: SMP-14-005.tex 304054 2015-09-20 16:15:07Z alverson $
\newlength\cmsFigWidth
\ifthenelse{\boolean{cms@external}}{\setlength\cmsFigWidth{0.85\columnwidth}}{\setlength\cmsFigWidth{0.4\textwidth}}
\ifthenelse{\boolean{cms@external}}{\providecommand{\cmsLeft}{top}}{\providecommand{\cmsLeft}{left}}
\ifthenelse{\boolean{cms@external}}{\providecommand{\cmsRight}{bottom}}{\providecommand{\cmsRight}{right}}

\newcommand{\zjets}{\ensuremath{\cPZ+\text{jets}}\xspace}
\newcommand{\zgjets}{\ensuremath{\cPZ/\gamma^{*}+\text{jets}}\xspace}
\newcommand{\vjets}{\ensuremath{V+\text{jets}}\xspace}
\newcommand{\gjets}{\ensuremath{\gamma+\text{jets}}\xspace}
\newcommand{\emu}{\ensuremath{\Pe\mu}\xspace}
\newcommand{\Zmumu}{\ensuremath{\cPZ\to\PGmp\PGmm}\xspace}
\newcommand{\Znunu}{\ensuremath{\cPZ\to \PGn\PAGn}\xspace}
\newcommand{\Zee}{\ensuremath{\cPZ\to\Pep\Pem}\xspace}
\newcommand{\ptZ}{\ensuremath{\pt^{\cPZ}}\xspace}
\newcommand{\ptG}{\ensuremath{\pt^{\gamma}}\xspace}
\newcommand{\njets}{\ensuremath{n_{\text{jets}}}\xspace}
\providecommand{\MADGRAPHPYTHIASIX}{\MADGRAPH{+}\PYTHIA{6}\xspace}
\providecommand{\BLACKHATSHERPA}{\BLACKHAT{+}\SHERPA\xspace}

\cmsNoteHeader{SMP-14-005}
\title{Comparison of the $\cPZ/\gamma^{*}+\text{jets}$ to $\gamma+\text{jets}$ cross sections in pp collisions at $\sqrt{s} = 8$\TeV}

\date{\today}

\abstract{
A comparison of the differential cross sections for the processes $\cPZ/\gamma^{*}+\text{jets}$ and photon $(\gamma)+\text{jets}$ is presented. The measurements are based on data collected with the CMS detector at $\sqrt{s}=8$\TeV corresponding to an integrated luminosity of 19.7\fbinv. The differential cross sections and their ratios are presented as functions of \pt. The measurements are also shown as functions of the jet multiplicity. Differential cross sections are obtained as functions of the ratio of the $\cPZ/\gamma^{*}$ \pt to the sum of all jet transverse momenta and of the ratio of the $\cPZ/\gamma^{*}$ \pt to the leading jet transverse momentum. The data are corrected for detector effects and are compared to simulations based on several QCD calculations.
}

\hypersetup{%
pdfauthor={CMS Collaboration},%
pdftitle={Comparison of the Z/gamma*+jets to gamma+jets cross sections in pp collisions at sqrt(s)= 8 TeV},%
pdfsubject={CMS},%
pdfkeywords={CMS, physics, Z, photons, ratio, jets, QCD, BSM}}

\maketitle
\section{Introduction}
\label{intro}

The associated production of a $\cPZ/\gamma^*$ or a $\gamma$ with one or more jets has been extensively
studied in proton-proton collisions at the CERN LHC, by both the CMS~\cite{CMSZat7TeV_jets,CMSZat7TeV_angVar,CMS7TeV_rapidity_Z_gamma_jet,CMSPHat7TeV,PhysRevD.91.052008} and ATLAS~\cite{AtlasZat7TeV,AtlasPHat7TeV} Collaborations. Precise measurements of these processes provide important tests of the Standard Model (SM) as well as crucial inputs in the determination of parton densities in the proton~\cite{StirlingPaperZGamma}. Such measurements can improve the validation and tuning of the models used in Monte Carlo~(MC) simulation. These processes are also important backgrounds in searches for new physics.

In the limit of high transverse momentum of the vector boson $V$ $\left(\pt^{V}\right)$ and at leading order (LO) in perturbative quantum chromodynamics (QCD), effects due to the mass of the $\cPZ$ boson ($m_\cPZ$) are small, and the cross section ratio of $\zjets$ to $\gjets$ as a function of $\pt^{V}$
is expected to become constant, reaching a plateau for $\pt^{V} \gtrsim 300\GeV$~\cite{StirlingPaperZGamma}. (In this paper, production of \zgjets~is denoted by \zjets.) A QCD calculation at next-to-leading order (NLO) for $\Pp\Pp\to\zjets$ and $\Pp\Pp\to\gjets$ was provided by the \BLACKHAT Collaboration~\cite{BlackHat}.
The NLO QCD corrections tend to lead to a decrease in the plateau value of the cross section ratio, while electroweak (EW) corrections are relatively small. However, at higher energies, EW corrections and QCD processes can introduce a dependence of the cross section on logarithmic terms of the form $\ln(\pt^{\cPZ}/m_{\cPZ})$ that can become large and pose a challenge for perturbative calculations such as \BLACKHATSHERPA~\cite{BlackHatNTuples}.
A precise measurement of the ($\Pp\Pp\to\zjets$)/($\Pp\Pp\to\gjets$)~cross section ratio provides important information about
the higher-order effects of these large logarithmic corrections at higher $\pt$.

In addition, searches for new particles involving final states characterized by the presence of large missing transverse energy ($\ETslash$) and hard jets, as described for example in Refs.~\cite{MT2} and~\cite{Chatrchyan:2014lfa}, use the $\gjets$ process to model the invisible $\cPZ$ decays, $\Znunu$, since the $\gjets$ cross section is larger than the $\zjets$ process where the $\cPZ$ decays to leptons. Measurements of the cross section ratio for $\zjets$ and $\gjets$ can help reduce uncertainties related to the $\Znunu$ background estimation in these searches.

We present precise measurements of both production cross sections and the cross section ratio for these two processes as a function of $\pt^V$. The results are compared with theoretical estimations.
The data sample was collected at the LHC during the 2012 run with the CMS detector in proton-proton collisions at a center-of-mass energy $\sqrt{s} = 8\TeV$ and corresponds to an integrated luminosity of 19.7\fbinv.
The $\cPZ$ bosons are identified via their decays to pairs of electrons or muons.
Measurements are made for different jet multiplicities ($n_{\text{jets}}\geq 1,\,2,\,3$) and for a subset requiring a large hadronic transverse energy ($\HT>300\GeV$, where $\HT$ is the scalar sum of all selected jet $\pt$ after jet identification). These requirements are meant to mimic the phase space requirements for analyses searching for new physics with an all-hadronic signature.

This paper is organized as follows: Section~\ref{cms} provides a brief description of the CMS detector; Section~\ref{mcprod} gives details of the Monte Carlo generators used in this analysis; Section~\ref{eventselection} describes the event selection; Section~\ref{resultZGammaJets} contains details about the background subtraction and the unfolding of the detector effects; Section~\ref{sec:systematics} discusses the sources of systematic uncertainties; Section~\ref{sec:results} presents the cross section and cross section ratio measurements for \zjets{} and \gjets{} production. Section~\ref{sec:conclusions} concludes with a summary of our results.

\section{CMS detector}
\label{cms}

The central feature of the CMS apparatus is a superconducting solenoid of 6\unit{m} internal dia\-meter, providing a magnetic field of 3.8\unit{T}.
The CMS experiment uses a right-handed coordinate system, with the origin at the nominal interaction
point, the $x$ axis pointing to the center of the LHC ring, the $y$ axis pointing up (perpendicular to
the LHC ring), and the $z$ axis along the anticlockwise-beam direction. The polar angle $\theta$ is
measured from the positive $z$ axis and the pseudorapidity is defined as $\eta= -\ln[\tan(\theta/2)]$.

Within the superconducting solenoid volume are a silicon pixel and strip tracker,
a lead tungstate crystal electromagnetic calorimeter (ECAL),
and a brass and scintillator hadron calorimeter. Each subdetector is composed of a barrel and two endcap sections.
Muons are measured in gas-ionization detectors embedded in the steel flux-return yoke outside the solenoid.
Extensive forward calorimetry complements the coverage provided by the barrel and endcap detectors.
In the barrel section of the ECAL, an energy resolution of about 1\% is achieved for unconverted or late-converting photons in the tens of \GeV energy range. The remaining barrel photons have a resolution of about 1.3\% up to a pseudorapidity of $\abs{\eta}=1$, rising to about 2.5\% at $\abs{\eta}=1.4$. In the endcaps, the resolution of unconverted or late-converting photons is about 2.5\%, while the remaining endcap photons have a resolution between 3 and 4\%~\cite{CMS-PAS-EGM-14-001}. The dielectron mass resolution for $\Z \to \Pe \Pe$ decays when both electrons are in the ECAL barrel is 1.8\%, and is 2.7\% when both electrons are in the endcaps. The electron momenta are estimated by combining energy measurements in the ECAL with momentum measurements in the tracker~\cite{CMS-PAS-EGM-10-004}. Muons are measured in the pseudorapidity range $\abs{\eta}< 2.4$, with detection planes made using three technologies: drift tubes, cathode strip chambers, and resistive-plate chambers. Matching muons to tracks measured in the silicon tracker results in a relative transverse momentum resolution for muons with $20 <\pt < 100\GeV$ of 1.3--2.0\% in the barrel and better than 6\% in the endcap. The \pt{} resolution in the barrel is better than 10\% for muons with \pt{} up to 1\TeV~\cite{Chatrchyan:2012xi}. A more detailed description of the CMS system can be found in Ref.~\cite{Chatrchyan:2008zzk}.

\section{Monte Carlo samples}
\label{mcprod}

Monte Carlo simulation samples are used to correct the data for acceptance and efficiency and for the \vjets~processes. They are also used to estimate the background to the \zjets~signal.

The \zjets~signal is generated with the \MADGRAPH (version 5.1.3.30)~\cite{MadGraph5} program. The leading-order multiparton matrix element (ME) calculation includes up to four partons (gluons and quarks) in the final state.
The showering and hadronization of the partons, as well as the underlying event, are modeled by \PYTHIA (version 6.4.26)~\cite{pythia6} with the $\cPZ2^{*}$ tune~\cite{TuneZ2star}.
The \kt MLM matching scheme~\cite{MatchingPaper} with a matching parameter of 20\GeV is applied to
avoid a double counting of final states arising in the ME calculation and the parton shower (PS).
The events are generated with the CTEQ6L1~\cite{CTEQ6} parton distribution functions (PDF) and rescaled using a global next-to-next-to-leading-order (NNLO) $K$-factor to match the inclusive cross section calculated with \FEWZ 3.1~\cite{FEWZ}. Backgrounds to \zjets~are generated using \MADGRAPH with the same configuration as the signal events. These include top quark-antiquark pairs (\ttbar) and EW backgrounds, such as $\PW+\text{jets}$ and diboson processes ($\PW\cPZ$, $\cPZ\cPZ$, $\PW\PW$).

The \zjets~signal and \ttbar background processes are also generated with \SHERPA (version 1.4.2) \cite{sherpa}, using the CT10 PDF~\cite{CT10}.
The cross section for the signal is also rescaled using a global next-to-next-to-leading-order (NNLO) $K$-factor.
For the \ttbar~background, the NNLO calculation provided by Ref.~\cite{ttbarXS} is used to calculate the NNLO $K$-factor. The background yields from EW processes are rescaled using the MCFM~\cite{MCFM} NLO cross sections.

In addition to these general purpose MC signal data sets, we use an NLO perturbative QCD calculation of \zjets~from the \BLACKHAT Collaboration~\cite{BlackHat}, which is available for a $\cPZ$ boson accompanied by up to three jets. These simulations use MSTW2008nlo68cl~\cite{MSTW} with $\alpha_{S}=0.119$ as the PDF set, and the renormalization and factorization scales ($\mu_{R}$ and $\mu_{F}$, respectively) are set to
\begin{equation*}
\mu_{R}=\mu_{F}=\HT^{p}+\ET^{\cPZ}\equiv\sum_{j} \pt^{j}+\sqrt{m_{\cPZ}^{2}+\left(\pt^{\cPZ}\right)^{2}},
\end{equation*}
where $\pt^{j}$ is the transverse momentum of the $j$th parton in the event and $\HT^{p}$ is the scalar $\pt$ sum of all outgoing partons with $\pt>20\GeV$. The CT10 and NNPDF2.3~\cite{NNPDF} PDF sets with $\alpha_{S}=0.119$ are used as a cross check and to estimate the theoretical systematic uncertainties.

The \BLACKHATSHERPA simulated events are organized into different types of processes to facilitate the calculation. An NLO estimation at $n$ jet level is obtained by combining tree-level (LO) calculations from the $n+1$ jet case to $n$ jet tree- and loop-level calculations. The Born and real emission calculations at both $n$ and $n+1$ jet levels are supplied by \SHERPA, while \BLACKHAT provides the NLO virtual loop-level correction terms. (For simplicity, \BLACKHATSHERPA will be referred to as \BLACKHAT.) The structure of the generated files and the preselections used in the simulation, as well as more details about \BLACKHAT, are described in Ref.~\cite{BlackHatNTuples}.

The \gjets~signal is simulated by \MADGRAPH, including up to four-parton final
states in addition to the photon. Fixed-order cross section calculations for \gjets~are affected by an instability due to dependencies on soft-gluon radiation, which can be overcome using all-order resummation~\cite{Frixione:1997ks,Banfi:2003jj}. The background contribution due to multijets is determined using control samples in data. The uncertainty in the photon purity is estimated with MC background samples simulated with \PYTHIA{6}. Multijet events in the \PYTHIA{6} sample with signal-like behavior can be enhanced by applying a filter that requires jet signatures with large electromagnetic deposits in the final state, e.g., jets with hadrons decaying into high-\pt photons. As an alternative method of estimating this background, we use a \MADGRAPH sample that includes jet production with as few as two and as many as four outgoing partons in the ME calculation.

We also simulate the \gjets~signal using \BLACKHAT. The overall procedure is analogous to the \zjets~\BLACKHAT samples, and $\gjets$ samples are available for $\gamma$ + 1, 2, and 3 jets. For \gjets, we use the following renormalization and factorization scales:
\begin{equation*}
\mu_{R}=\mu_{F}=\HT^{p}+\ET^{\gamma}\equiv\sum_{j} \pt^{j}+\pt^{\gamma}.
\end{equation*}

The \BLACKHAT production requires that the photons satisfy the Frixione cone isolation condition~\cite{Frixione}
\begin{equation*}
\sum_{i}\ET^i~\Theta\left(\delta-R_{i\gamma}\right)\leq\mathcal{H}\left(\delta\right),
\end{equation*}

for all $\delta$ less than $\delta_{0}$ around the axis of the photon. Here, $R_{i\gamma}$ is the distance in $\eta$ and azimuthal angle $\phi$ between the $i^{\text{th}}$ parton and the photon, and $\Theta$ is the step function. The function $\mathcal{H}\left(\delta\right)$ is chosen such that it vanishes as $\delta\to0$. In particular,
\begin{equation*}
\mathcal{H}\left(\delta\right)=\ET^{\gamma}~\epsilon\left(\frac{1-\cos\delta}{1-\cos\delta_{0}}\right)^n.
\end{equation*}

The Frixione cone in effect only adds contributions from partons which are within $\delta_{0}$ of the photon. In the \BLACKHAT samples,  $\epsilon=0.025$, $\delta_{0}=0.4$, and $n=2$. These were selected because a Frixione cone with these choices mimics the selections in the true on-shell photon definition at particle level. Photon distributions using Frixione cone requirements are found to agree with those using cone isolation to within 1--2\%~\cite{Bern:2011pa}.

The simulation of the CMS detector is based on the \GEANTfour package~\cite{geant4}.
The simulated events used for the detector level MC estimations are reconstructed following the same procedures used for the data.
For our run, the average number of inelastic proton-proton collisions occurring per LHC bunch crossing was 21. The correct distribution of the number of pileup events overlapping the hard interaction process per bunch crossing is taken into account in the MC by reweighting the simulated minimum bias events to match the spectrum of pileup interactions observed in data.

\section{Event selection and object reconstruction}
\label{eventselection}

The selection of $\zjets$ events begins by requiring
two same-flavor high-\pt leptons (electrons or muons) at trigger level. The $\pt$ threshold of the trigger objects
is 17\GeV for the leading muon (the muon with the largest $\pt$) and 8\GeV for the subleading muon.
The dielectron trigger requires the same thresholds of 17 and 8\GeV on the $\pt$ of the leading and subleading electron candidates, respectively.
Additionally, the trigger requires that the electron candidates be isolated from other energy
deposits in the calorimeter so an isolation requirement is imposed on the electron track.

Muons are reconstructed offline by a simultaneous fit of hits recorded in the silicon tracker
and in the muon detectors~\cite{Chatrchyan:2012xi}. Electrons are reconstructed from energy clusters
in the ECAL and tracking information~\cite{CMS-PAS-EGM-10-004}.
The two leading leptons are required to be of opposite electric charges and of the same flavor, with $\pt>20\GeV$ and $\abs{\eta}<2.4$. For both candidates, a match with a corresponding trigger object is required.
The dilepton invariant mass, $m_{\ell\ell}$, is required to satisfy $71\GeV<m_{\ell\ell}<111\GeV$. This will be referred to as the ``\cPZ~boson mass window''.

The particles in the event are reconstructed using the particle-flow (PF) technique~\cite{CMS-PAS-PFT-09-001,CMS-PAS-PFT-10-001}, which consists of identifying each single particle with an optimized combination of all subdetector information. Depending on their signatures in the various subdetectors, particles fall into five different PF categories: muons, electrons, photons, neutral hadrons, and charged hadrons. The lepton candidates are required to be isolated from the other particles in the event, so to evaluate the isolation a scalar $\pt$ sum of PF objects is calculated in the cone $\Delta R=\sqrt{\smash[b]{(\Delta\phi)^2+(\Delta\eta)^2}}$ around the direction of the object. The contribution from pileup to this isolation scalar \pt sum is subtracted using
the average pileup energy per unit area in the $\eta$-$\phi$ plane evaluated for each event~\cite{Cacciari:JetArea}.
For electrons, the pileup-subtracted isolation sum is calculated in a cone of
$\Delta R=0.3$ around the direction of the electron and is required to be below 15\% of the electron \pt. For muons, the radius
is set to be $\Delta R=0.4$ and the isolation variable is required to be less than 12\% of the muon $\pt$. Lepton reconstruction, identification, and isolation efficiencies are measured using the ``tag-and-probe'' technique as described in Ref.~\cite{inclusiveWZ}. Efficiencies for simulated events are corrected using $\eta$- and \pt-dependent scale factors to account for differences between data and simulation.
Scale factors typically range between 0.98 and 1.02.

The photons are reconstructed offline from energy clusters in ECAL~\cite{CMS-PAS-EGM-14-001}. Events for the $\gjets$ processes are selected at the trigger level, where the presence of a high-\pt photon candidate is required.
Since the instantaneous luminosity of the LHC increased during the
data-taking period, the threshold in $\pt$ increased as well, and the lower $\pt$ threshold triggers are prescaled in order to keep the rate at a reasonable level.  An unprescaled trigger is available only for
a photon with transverse momentum $\pt^{\gamma}> 207\GeV$. In order to further reduce the rate,
a loose shower shape cutoff $\sigma_{\eta\eta}<0.24$ is imposed at trigger level,
where $\sigma_{\eta\eta}$ measures the extension of the shower in pseudorapidity in terms of the energy-weighted spread within the 5$\times$5 crystal matrix around the most energetic crystal in the photon cluster. For photon candidates, a match with a corresponding trigger object is required. For this analysis, only isolated high-\pt photons located inside the barrel region
of the detector ($\abs{\eta}<1.4$) are considered. We concentrate on photons inside the barrel region because the data size of the templates, described in Section~\ref{gjetsselection}, allows for a precise purity determination in this region.
Around 40\% of the photons convert into $\Pep\Pem$ pairs inside the tracker material.
Conversion track candidates are fitted from a combination of ECAL seeded tracks
and Gaussian sum filter~\cite{GaussianSum} electron tracks originate from a common vertex.
The track pair is then matched to energy clusters in ECAL to identify
a converted photon candidate.
The final photon candidates are checked for possible overlap with electron candidates by looking for electron track seeds in the pixel detector or by using the characteristics of the track pair for converted photons. Isolation requirements are
separately imposed on the pileup-corrected scalar \pt sum of neutral and charged hadrons, as well as on additional photons inside a cone of $\Delta R=0.3$ around the photon candidate direction. MC over data scale factors for the selection efficiencies of unconverted and converted photons without
the electron veto are measured
using the tag-and-probe technique on \Zee~events; the scale factors for the electron veto
efficiency on signal photons are determined using $\cPZ\to\mu^+\mu^-\gamma$ candidates. These scale factors range between 0.96 and 1.01 for photon candidates with $\pt^{\gamma}>40\GeV$.

Jets are reconstructed from the four-momentum vectors of all PF objects. The anti-\kt clustering algorithm~\cite{Cacciari:2008antikt} is used here with a distance parameter of $R=0.5$ in its \FASTJET~\cite{fastjet2} implementation. The jets are clustered by four-momentum summation. The reconstructed PF candidates are calibrated separately to account for the nonlinear and
nonuniform response of the CMS hadron calorimeter, especially for neutral hadrons.
Charged hadrons and photons are well measured in the silicon tracker and the ECAL, and therefore need only minimal corrections.
Thus, the resulting jets require only small additional momentum adjustments. Jet energy corrections are obtained
using \GEANTfour simulated events generated with \PYTHIA{6}. The energy contributions due to the presence of additional proton-proton interactions are subtracted from each jet using the measured pileup unit density in the event and the jet area~\cite{Cacciari:JetArea}. The $\eta$ dependent corrections are adjusted using exclusive dijet events, while the \pt-dependent corrections are adjusted using exclusive $\Z+1\text{-jet}$ and $\gamma+\text{1-jet}$ events in the data~\cite{CMS-JME-10-011}.
As a result of these adjustments, the reconstructed jets are corrected to the stable particle level~\cite{Buttar:2008jx}.
For PF jets, the jet energy correction factor typically ranges from 1 to 1.2.
Jets originating from pileup are rejected using the criteria described in Ref.~\cite{CMS-PAS-JME-13-005}. This rejects 90--95\% of pure pileup jets while keeping over 99\% of jets from the primary interaction. Jet identification quality requirements are imposed in order to remove spurious jets caused by noise in the calorimeter. The remaining jets are accepted for the analysis if they satisfy $\pt>30\GeV$ and $\abs{\eta}<2.4$.
Additionally, jets within a radius of $\Delta R<0.5$ with respect to the axes of each lepton or photon candidate are removed. This cut affects a small number of jets. For both $\zjets$ and $\gjets$ selections,
the presence of at least one jet is required.

The selection of $\zjets$ events is separate from the selection of $\gjets$ events, and the two data samples
are analyzed and corrected independently. The overlap between $\zjets$ and $\gjets$ events is negligible. The analysis is repeated in four different, but not mutually exclusive, kinematic regions, with $\pt^{V}>100$\GeV and
\begin{itemize}
\item $n_{\text{jets}}\geq1$,
\item $n_{\text{jets}}\geq2$,
\item $n_{\text{jets}}\geq3$,
\item $\HT>300\GeV$.
\end{itemize}
The rapidity of the $\cPZ$ boson is not restricted for the individual $\cPZ$ boson distributions.
However, it is restricted to the rapidity range $|y|<1.4$ for the distributions of the ratio of $\ptZ$ to $\ptG$ because the photon is measured only in this central rapidity range. Rapidity is defined as $y = \frac{1}{2}\ln\left[\left(E+p_{z}\right)/\left(E-p_{z}\right)\right]$. The measured differential cross sections are binned in equal intervals of $\log_{10}\pt$(\GeV) of width 0.045 from 100 to 800\GeV, corresponding to the overlap region
between the \zjets~and \gjets~phase space. This binning ensures that as the number of events decreases, the bin width increases in a regular way. In terms of the photon purity determination (defined in Section~\ref{gjetsselection}), the bins are chosen such that there are enough events in all bins in the final distribution to ensure a reliable measurement.

\section{\texorpdfstring{Background determination and unfolding}{Background determination and unfolding}}
\label{resultZGammaJets}

\subsection{\texorpdfstring{The $\zjets$ selection}{The Z+jets selection}}

Events from the $\zjets$ process are selected as \Zee~and \Zmumu~candidates with one or more jets,
as described in Section \ref{eventselection}. The background-subtracted distributions are unfolded
to the stable particle level for each decay channel separately and then combined.

Several SM processes contribute to backgrounds to the $\zjets$ signal.
For low $\ptZ$, the most important background is \ttbar production, whereas at higher $\ptZ$ values, diboson production is the dominant background. Contributions due to
$\PW+\text{jets}$ and $\PW\PW+\text{jets}$ are negligible for this analysis.
The background contributions are subtracted using relative event rates predicted by \MADGRAPH after an NNLO scaling for Drell--Yan and \ttbar samples and an NLO scaling for the electroweak backgrounds.

A cross check of the validity of the procedure for \ttbar background estimation is performed using an \emu~control sample in data.
This sample is largely dominated by \ttbar production with an additional contribution from fully leptonic $\cPZ\to \tau^{+}\tau^{-}$ decays.
Both in
absolute scale and shape, the simulation reproduces the dilepton transverse momentum spectrum ($\pt^{\ell\ell}$) in the data within 10\%. This statement is valid both for a selection
with a relaxed dilepton mass of $m_{\emu}>60\GeV$ and a selection within the $\cPZ$ boson mass window as used for the final event selection.
As a second check, the relative rate of \emu~events in data and MC are compared to those of dielectron or dimuon events as a function of the dilepton $\pt^{\ell\ell}$.
Events with \emu~are selected by requiring the \emu~invariant mass to be either in the $\cPZ$ boson mass window or in
the whole mass range.
Events from $\Pep\Pem$ or $\PGmp\PGmm$ are selected in  the $\cPZ$ boson mass window.
All four distributions of these relative event rates from simulation are compatible with data within 10\%.
The \ttbar background peaks at around $\pt^{\ell\ell}\approx 100\GeV$, where it amounts to 1.5\% for the inclusive 1-jet selection and to 8\% for the inclusive 3-jet selection.
In the high-\HT selection, it amounts to up to 12\%.
For $\HT>300\GeV$, the relative rate drops below 0.5\%, while in the \emu~channel no event is observed
beyond $\pt^{\emu}>450\GeV$.

At around $\pt^{\ell\ell}\approx 150\GeV$, the EW background increases and reaches a plateau of about 5--7\%
for all phase space selections of the analysis beyond $\pt^{\ell\ell}\approx 400\GeV$.
The rate of the combined EW backgrounds predicted by simulation in a control region of multilepton final states is checked with data in the following way.
Instead of selecting the two leading leptons, and then enforcing the same flavor requirement, we instead select the first two leptons matching the trigger objects with the same flavor. The rate of events gained with respect to the baseline selection is largely dominated by diboson events.
These additional data events are compared to
estimations from \MADGRAPH, finding an agreement
within roughly 10\% for all jet multiplicity phase space selections.
This comparison is done in the range $\pt^{\ell\ell}<300\GeV$.

 The same selection criteria from data are used at the particle level: leading leptons are required to have $\pt>20\GeV$ and $\abs{\eta}<2.4$,
while jets are required to have $\pt>30\GeV$
within the region of $\abs{\eta}<2.4$. The particle
level jets in simulation are obtained by clustering the generated stable particles (after hadronization and including neutrinos) using the anti-\kt algorithm
with distance parameter of $R=0.5$.
Electrons and muons have different energy losses due to final state radiation at particle level. In order to compensate for these differences, we define a ``dressed'' level to make the electron and muon channels compatible to within 1\%. This is achieved by defining in simulation a particle momentum vector by adding the momentum of the stable lepton and the momenta of all photons with a radius of $\Delta R=0.1$ around the stable lepton.
All jets are required to be separated from each lepton by $\Delta R>0.5$.

The background-subtracted detector-level distributions from data are unfolded to the particle level.
The unfolding response matrix includes detector resolution effects and efficiencies.
We use \MADGRAPH to build a response matrix which allows us to map detector-level distributions to particle level. To quantify the bias introduced by the choice of the MC model, we use \SHERPA as an alternative.
The off-diagonal elements of the response matrices are small for both channels. For the dielectron channel, 85--95\% of all events in a given bin of the reconstructed $\pt^{\mathrm{ee}}$ distribution are mapped onto the same bin at the particle level.
For the dimuon channel, at low $\pt^{\mu\mu}$ around 85\% fall in this category,
whereas at very high $\pt^{\mu\mu}$, only 67\% stay in the same bin at the particle level.
The remaining events typically fluctuate to directly neighboring bins.
The iterative method used by d'Agostini~\cite{dAgostini}, as implemented in the \textsc{RooUnfold} package~\cite{RooUnfold}, is used to regularize the inversion of the matrix.
Subsequently, the unfolded distributions from \Zee~and \Zmumu~are found to be compatible. They are combined using the best linear unbiased estimator~\cite{BLUE} to obtain the final distributions. The resulting averaged leptonic \zjets~distributions from both channels are not corrected to the total cross section.

\subsection{The \texorpdfstring{$\gjets$}{g+jets} selection}
\label{gjetsselection}

After selecting \gjets~events (as described in Section~\ref{eventselection}),
the photon signal purity is determined in each \pt bin.
The main background is due to QCD multijet production,
where either one of the jets, or an electron or $\Pgpz$ inside a jet, is misidentified as a photon candidate.
Since simulations do not provide a reliable description of this background,
the purity, which is defined as the number of true isolated photons from the hard scattering versus the number of all photon candidates, is determined from data.
At the particle level, a true isolated photon is defined as a prompt photon,
 around which the scalar sum of the $\pt$ of all stable
particles in a cone of radius $\Delta R=0.4$ is less than 5\GeV.
Similarly, at the detector level, for each \pt bin of the photon spectrum, the purity is determined through a fit of
the photon isolation sum variable $I_\text{ph}^\mathrm{PF}$, defined
as the scalar $\pt$ sum of all other PF photons around the axis of the selected photon candidate,
inside a cone of $\Delta R=0.4$.
The sum is corrected for the pileup contribution and
the energy deposit (``footprint'') of the selected photon candidate itself.
The $I_\text{ph}^\mathrm{PF}$ distribution for the data is fitted as a sum of signal and background template distributions,
in each \pt bin, in order to calculate the purity $f$:
	$I_\text{ph}^\mathrm{PF}(\text{data}) = f \, I_\text{ph}^\mathrm{PF}(\text{signal}) +
	(1-f) \, I_\text{ph}^\mathrm{PF}(\text{background})$.

\begin{figure}[hbtp]
  \centering
    \begin{minipage}[c]{0.50\textwidth}
    \centering
    \includegraphics[width=1.0\textwidth]{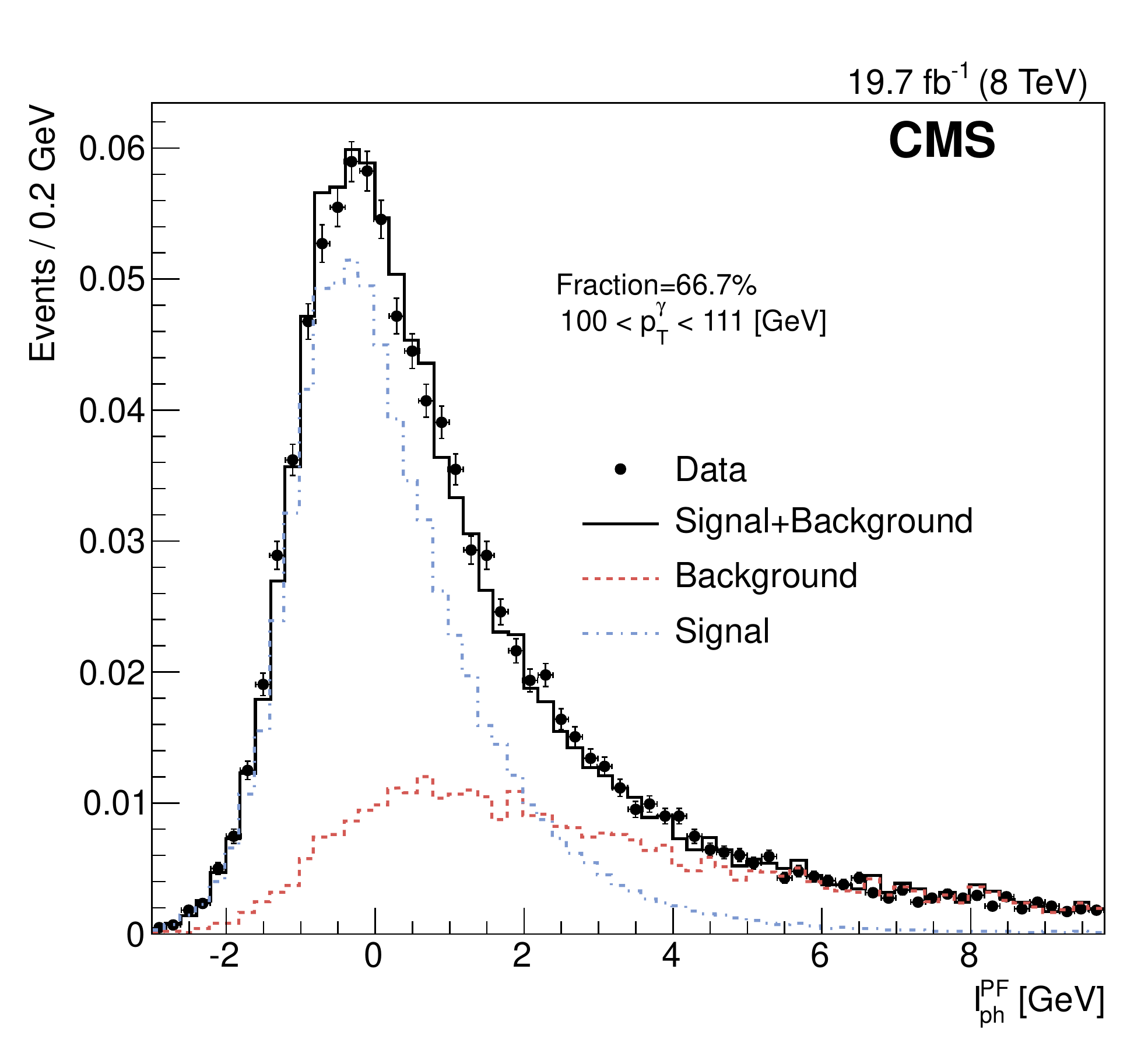}
  \end{minipage}%
  \begin{minipage}[c]{0.50\textwidth}
    \centering
    \includegraphics[width=1.0\textwidth]{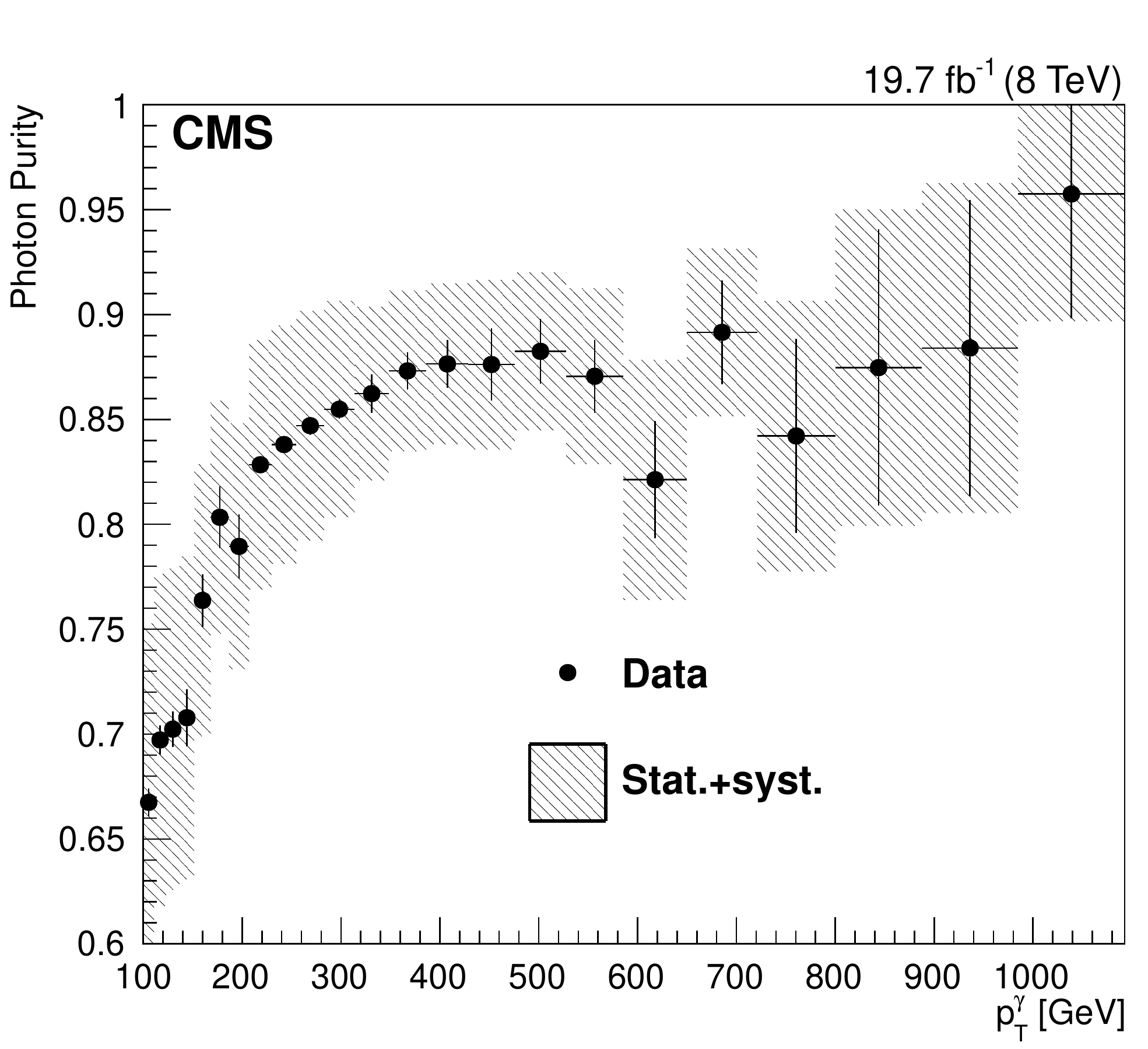}
  \end{minipage}
  \caption{The purity fit on the photon component of the photon isolation $I_\text{ph}^\mathrm{PF}$ in the photon transverse momentum bin between 100 and 111\GeV in data (left). The photon purity as function of the photon transverse momentum (right). The dots are the data points, the dot-dashed line is the signal template, and the dotted line represents the background component. The solid line represents the fit and the legend shows the resulting purity fraction of 66.7\%.}
  \label{fig:photon_purity_template}
\end{figure}

In order to model the contribution of the underlying event to the photon component isolation sum
around the signal photon candidate, a signal template is obtained from the data
through the random-cone (RC) method~\cite{CMS-PAS-SMP-13-001}.
After selecting photon candidates fulfilling a requirement on the  shower shape of $\sigma_{\eta\eta}<0.011$,
a cone of $\Delta R =0.4$ is randomly chosen in $\phi$
at the same $\eta$ as the photon candidate, excluding the back-to-back direction to avoid selecting any
recoiling jet.
The candidate cone is rejected if it contains objects originating from a hard interaction,
e.g., jets with $\pt>30\GeV$ or photons with $\pt>20\GeV$.
The RC templates show a very good agreement over orders of magnitude between data, simulation, and
the true-photon MC templates obtained by matching a detector-level photon candidate
with an isolated photon at the particle level.
Background templates are
constructed by selecting photon candidates in the data with an inverted shower shape requirement, $0.011<\sigma_{\eta\eta}<0.014$. Since there are a small number of background events with high $\pt$ photon candidates, the templates are obtained in wider bins of $\pt^{\gamma}$ than used in the final analysis.
After construction of the templates, a binned maximum-likelihood fit to the $I_\text{ph}^\mathrm{PF}$ data distributions is performed as a sum of the signal and background template distributions.
The statistical uncertainty on the fit includes the effect of the limited template sample size. An example of the fit for the photon component of the photon isolation can be seen in Fig. \ref{fig:photon_purity_template} (left). The results of such fits are displayed in Fig. \ref{fig:photon_purity_template} (right), where the black dots represent the measured purity and the solid lines represent the statistical uncertainties including the template uncertainties. The purity ranges from around 65\% in the lower $\pt$ bins to 90\% at high $\pt$. The shaded band represents the total statistical and systematic uncertainties in the purity measurement. The difference in the fractions obtained from the closure test on simulation is treated as the systematic uncertainty.
True MC signal (background) templates are determined using identified photons at detector level matched (not matched) to a particle-level photon.
The closure test uses the same approach for deriving templates that was used in data: it takes simulated samples and compares the resulting templates of the data-driven approach with the templates constructed exploiting the MC-truth information. This is the dominant source of systematic uncertainty in the purity estimate.
Other effects, such as a change in the $\sigma_{\eta\eta}$ requirement, are found to be negligible.
The systematic uncertainties of the purity estimate are discussed further in Section~\ref{sec:systematics}. The data yields in each \pt bin, after correcting for purity, are unfolded to the particle level
with a procedure identical to the one used for the $\zjets$ process. Over the whole $\pt^{\gamma}$ spectrum, the diagonal elements of the response matrix contain more than 90\% of the events.

\section{Systematic uncertainties}
\label{sec:systematics}

For \zjets~and \gjets~we consider the following uncertainties: the jet energy scale (JES) uncertainty, the jet energy resolution (JER) uncertainty, uncertainties due to the MC model from the unfolding procedure (UF), the pileup uncertainty (PU), and the luminosity uncertainty (Lumi). Systematic effects specific to \gjets~are those related to the photon energy scale ($\gamma$ ES) and the purity determination ($\gamma$ Pur). For \zjets, we consider the background subtraction (BG) and the lepton (muon and electron) momentum scale (LS) and resolution (LRES) uncertainties, as well as uncertainties in the lepton efficiency and isolation. We also consider lepton efficiency scale factors (lep SFs) for the \zjets~events. For both processes, the uncertainty associated with the luminosity is a flat 2.6\% over the whole range~\cite{CMS-PAS-LUM-13-001}.

The systematic effects due to scale uncertainties affect the data, so we vary the momenta of the jets or the leptons independently within their uncertainties and rerun the unfolding on the shifted distribution. The differences in the final results are taken to be the systematic uncertainties.

The uncertainty due to the JES affects distributions through the jet $\pt$ threshold~\cite{CMS-JME-10-011}. For central jets, the JES uncertainty is around 3\% at 30\GeV, decreasing to 1\% at 100\GeV. Therefore, in the $\njets=1$ case, the boson $\pt$ is almost completely unaffected. For the 2- or 3-jet inclusive phase space selection, the requirement of additional jet activity increases the uncertainty in the photon and $\Z$ spectra due to JES to 5--10\% over the whole $\pt^{V}$ range. In the $\HT>300\GeV$ selection, the JES uncertainty is around 5--7\% at low $\pt^{V}$  and below 1\% for $\pt^{V}>400\GeV$.

Systematic effects due to purity and background subtraction have to be applied prior to unfolding to evaluate the uncertainty. The other uncertainties affect the response matrix in the simulation, and the unfolding is performed with these modified matrices to determine the relative uncertainty. For example, we modify the resolution of jets in the MC and then calculate a new response matrix with these modified resolutions. The difference between this result and the nominal result is taken as the uncertainty due to JER.

For the $\zjets$ process, the dominant sources of uncertainty are the lepton SFs and the LS in the $\njets\geq1$ case and the JES uncertainty otherwise. The uncertainty due to the background subtraction is typically below 1\%. The lepton resolution uncertainty has an effect that is typically less than 0.5\%.
The effect of the electron energy scale uncertainty increases with $\pt^{\cPZ}$ from 1\% at 40\GeV to 5\% at 800\GeV. For muons, the scale uncertainty has an effect ${<}1$\% up to 250\GeV, which increases up to 15\% at high $\pt^{\cPZ}$.
Above 200\GeV, the track becomes very straight, and so the influence of the muon system becomes more relevant with respect to the tracker for the muon $\pt$ distribution. This leads to an increase in the muon scale uncertainty.

For the unfolding procedure, an additional check using the matrix obtained from \SHERPA instead of that from \MADGRAPH is performed, resulting in a cross section uncertainty of 2--3\% for all phase space regions. A cross check using the Singular Value Decomposition regularization method~\cite{SVD} for the unfolding shows negligible deviations.

The JER is measured to be about 5\% larger than predicted in simulation for the central detector part
($\abs{\eta^\text{jet}}<1.4$) with an uncertainty of about 5\%, and roughly 10\% larger in the endcaps with an uncertainty of roughly 7\%~\cite{CMS-JME-10-011}.  The JER and pileup uncertainties in $\zjets$ events have values typically below 1\%. The uncertainties on \zjets~are summarized in Table \ref{tab:ZPT_Uncertainties}.

Aside from the $\pt$ spectrum, we also consider two additional variables for the \zjets~final state: the ratio of the $\ptZ$ to $\HT$ and to the $\log_{10}\left(\pt^\cPZ/\pt^{\mathrm{j}1}\right)$, where $\pt^{\mathrm{j}1}$ is the transverse momentum of the largest jet in the event. Most uncertainties in these distributions are similar to those described above for the $\ptZ$ spectra, with the exception of the JES uncertainty. The latter has a larger influence on hadronic quantities ($\HT$, $\pt$ of the jets, and \njets) which enter the distributions directly, rather than through phase space selections.

\begin{table}[htbp!]
\centering
\topcaption{Systematic Uncertainties for the $\ptZ$ Spectrum}
\begin{tabular}[h!]{l|ccccccccc}
\hline
Process&JES&JES&JER&Lep SFs&UF&PU, BG&LS&Lumi\\
&($\njets\geq1$)&(otherwise)&&&&LRES&&\\
\hline
$\Zee$&1--3\%&5--10\%&${<}1\%$&3--4\%&2--3\%&${<}1$\%&1--5\%&2.6\%\\
$\Zmumu$&1--3\%&5--10\%&${<}1\%$&2.5--5.5\%&2--3\%&${<}1$\%&${<}1\%$&2.6\%\\
\hline
\end{tabular}
\label{tab:ZPT_Uncertainties}
\end{table}

For the \gjets~process, the dominant uncertainty is due to the photon purity. This is a result of the difference between the shapes of the templates defined using the data-driven techniques from above and the distributions of the ``true'' templates for isolated photons in simulated events. Data samples are generated using the distributions of isolation variables for every bin of each variable with the fractions measured in data. Each of these is fitted with templates built in MC using the same techniques as on data, and the average difference between these fitted fractions and the generated fractions is quoted as the systematic uncertainty due to photon purity estimation~\cite{CMS-PAS-SMP-13-001}. This difference is around 10\% when $\pt^{\gamma}\approx 100\GeV$ and it decreases to roughly 4\% at $\pt^{\gamma}\approx 400\GeV$ for the inclusive $\njets\geq1$ selection. A change in the selection criteria on $\sigma_{\eta\eta}$ leads to negligible effects on the purity estimation. The background templates do not show any dependence on \HT or on the number of jets in the analysis.
Therefore, the same background templates of the inclusive selection are used for all phase space regions and a similar template uncertainty is obtained.

The JER uncertainty has
a negligible effect in the analysis region $\pt^{\gamma}>100\GeV$.
For the 2- or 3-jet phase space and high $\HT>300\GeV$
selection, the resolution uncertainty has an effect around 0.5--1.5\%. The effect of the $\gamma$ ES uncertainty on the cross section measurement is constant across the whole range and less than 3\%. The unfolding uncertainty is estimated by using an unfolding matrix from \PYTHIA{6} simulation and is around 2\%. The uncertainty in the pileup interactions is evaluated by rescaling the cross section of minimum bias events by 5\% in the MC reweighting procedure.
Typically, these uncertainties are very small, below 0.5\%. These systematic uncertainties for the $\gjets$ process are summarized in Table~\ref{tab:GPT_Uncertainties}.

\begin{table}[htbp!]
\centering
\topcaption{Systematic Uncertainties for the $\ptG$ Spectrum}
\begin{tabular}[h!]{l|ccccccccc}
\hline
Process&JES&JES&JER&UF&PU&$\gamma$ Pur&$\gamma$ ES&Lumi\\
&($\njets\geq1$)&(otherwise)&&&&&& \\
\hline
$\gamma$&1--3\%&5--10\%&0.5--1.5\%&2\%&${<}0.5\%$&4--10\%&3\%&2.6\%\\
\hline
\end{tabular}
\label{tab:GPT_Uncertainties}
\end{table}

\section{Results}
\label{sec:results}
\subsection{Differential cross sections}
\label{diffcrosssections}

\begin{figure}[hbtp]
  \centering
    \includegraphics[width=0.49\textwidth]{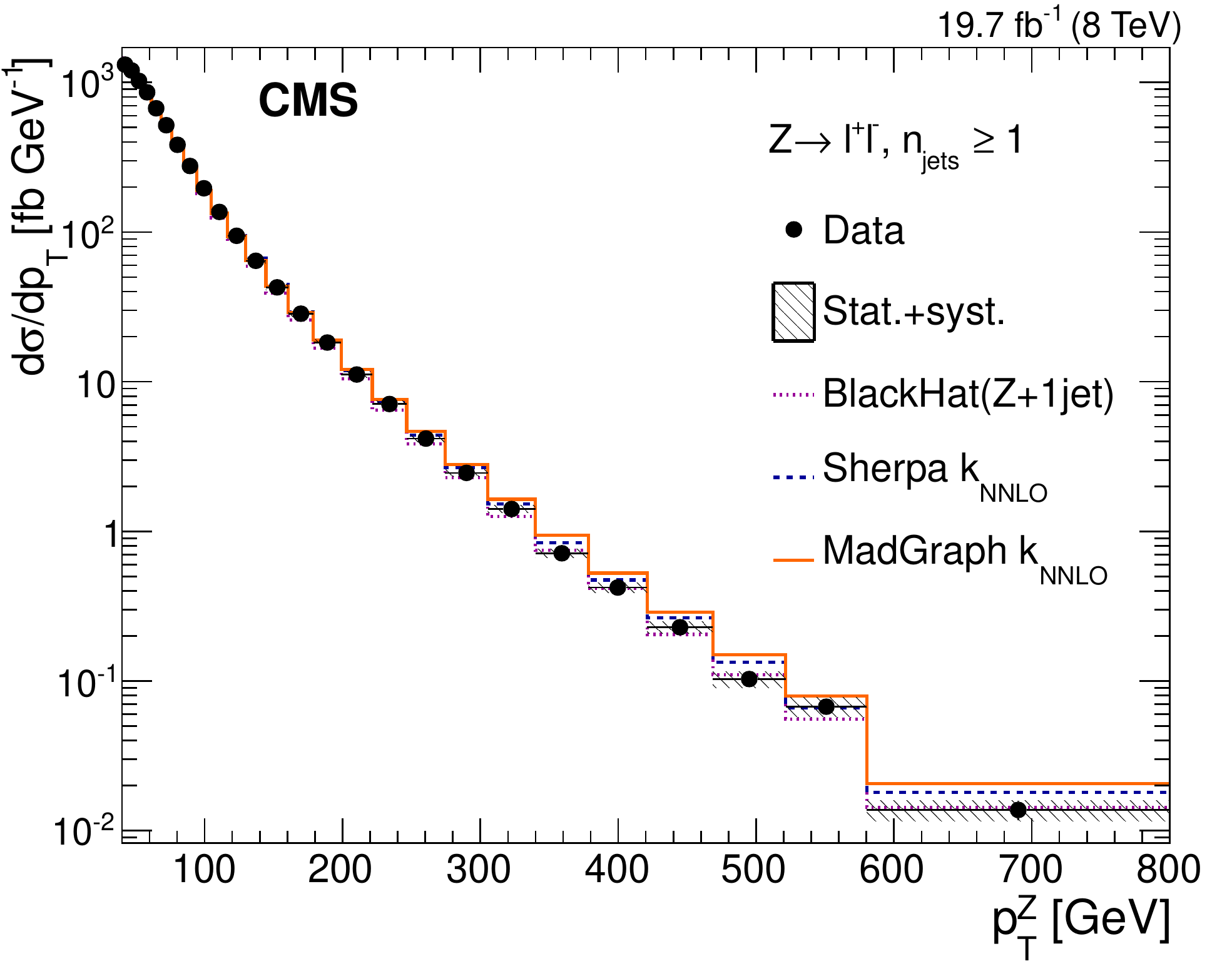}
    \includegraphics[width=0.49\textwidth]{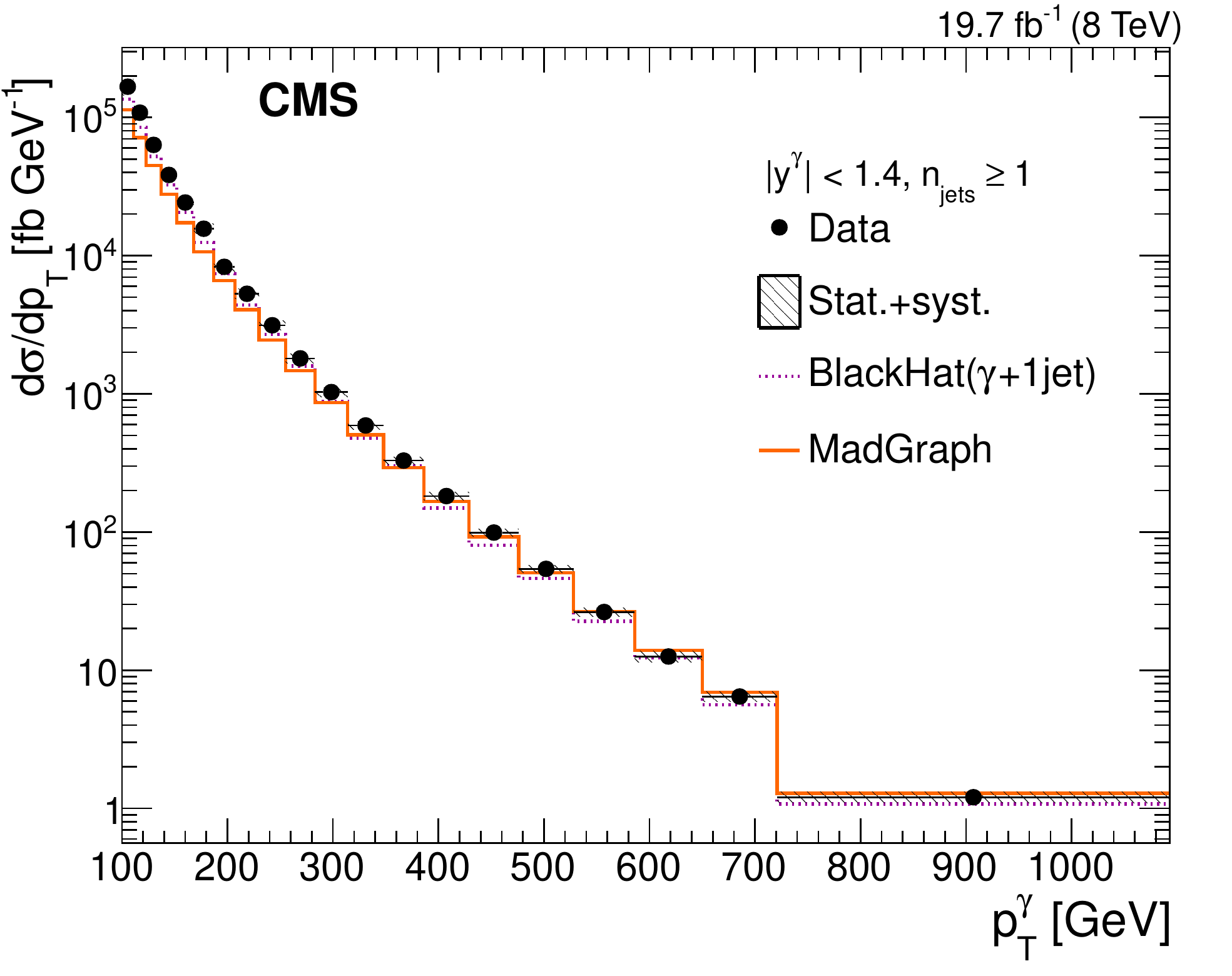}
  \begin{minipage}[c]{0.49\textwidth}
    \centering
    \includegraphics[width=1.0\textwidth]{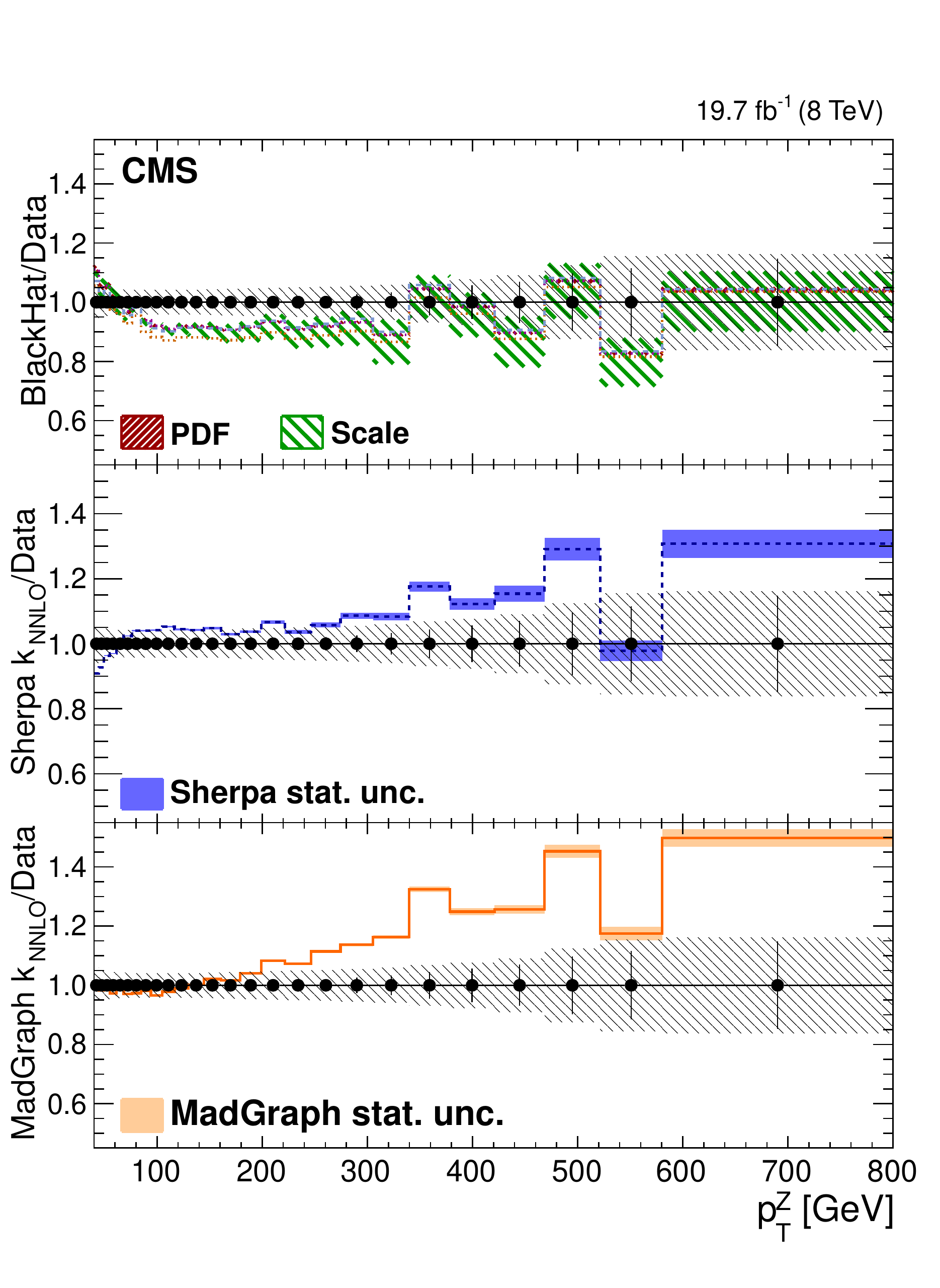}
  \end{minipage}%
  \begin{minipage}[c]{0.49\textwidth}
    \centering
    \includegraphics[width=1.0\textwidth]{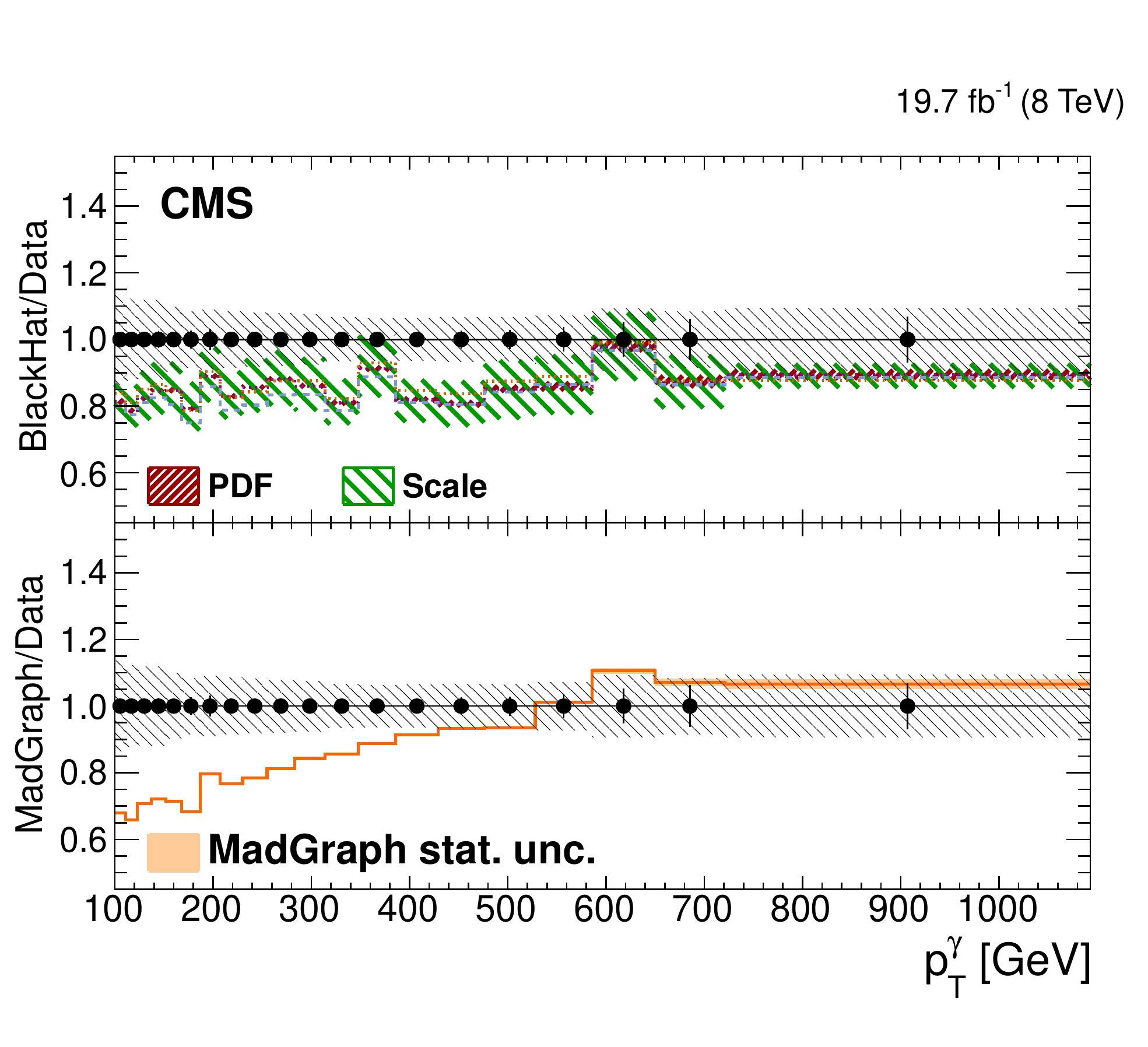}
  \end{minipage}
  \caption{Top left: Differential cross section for \cPZ~boson production as a function of $\ptZ$ for an inclusive \zjets, $\njets\geq1$ selection of detector-corrected data in comparison with estimations from \MADGRAPHPYTHIASIX, \SHERPA, and \BLACKHAT.
Top right: Differential cross section for photon production as a function of $\pt^{\gamma}$ for an inclusive \gjets, \mbox{$\njets\geq1$} selection for central rapidities $\abs{y^{\gamma}}<1.4$ in detector-corrected data is compared with estimations from \MADGRAPHPYTHIASIX and \BLACKHAT. A detailed explanation is given in Section~\ref{diffcrosssections}. The bottom plots give the ratio of the various theoretical estimations to the data in the \zjets~case (bottom left) and \gjets~case (bottom right).}
  \label{fig:pt_Bosons}
\end{figure}

\begin{figure}[hbtp]
  \centering
    \includegraphics[width=0.49\textwidth]{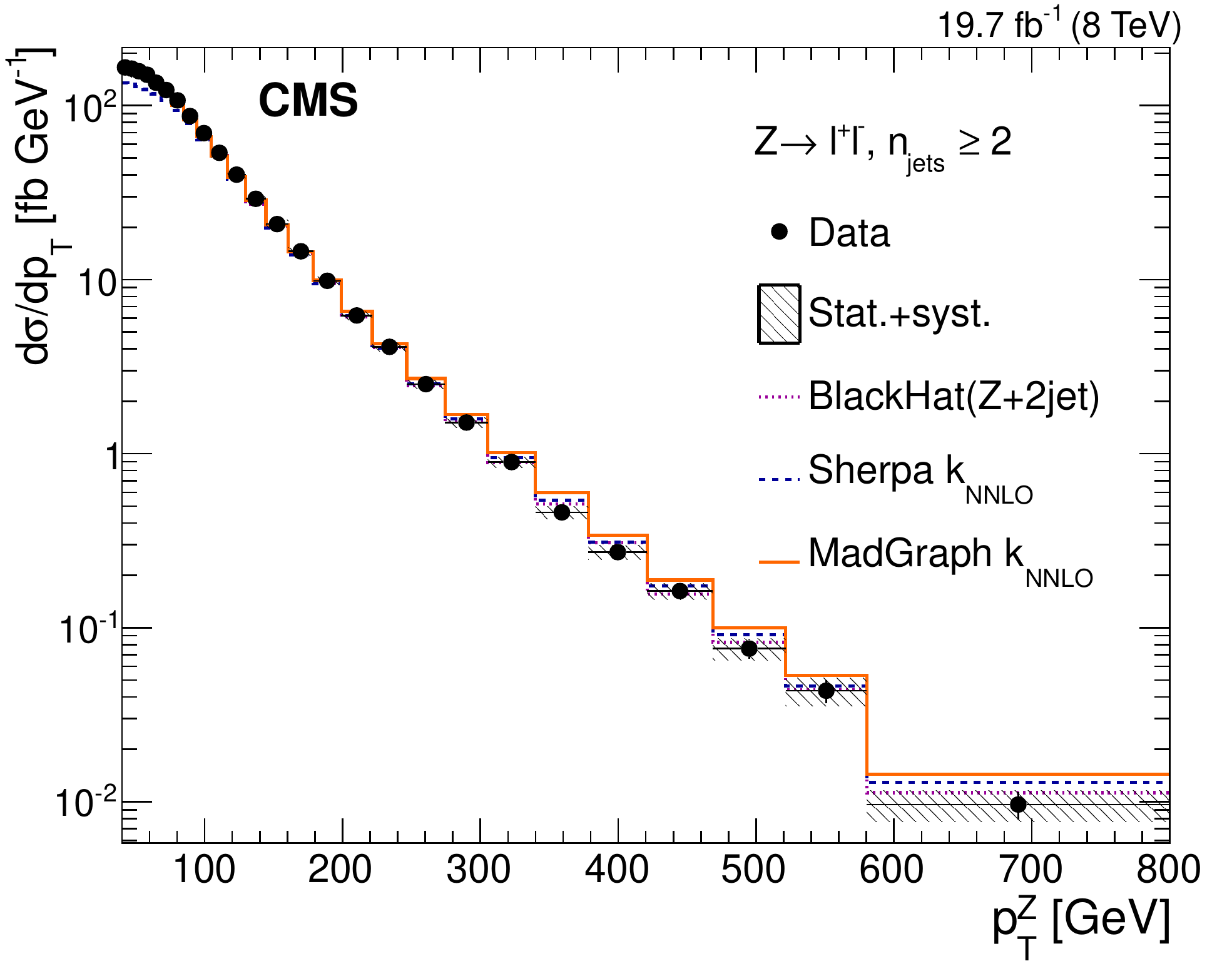}
    \includegraphics[width=0.49\textwidth]{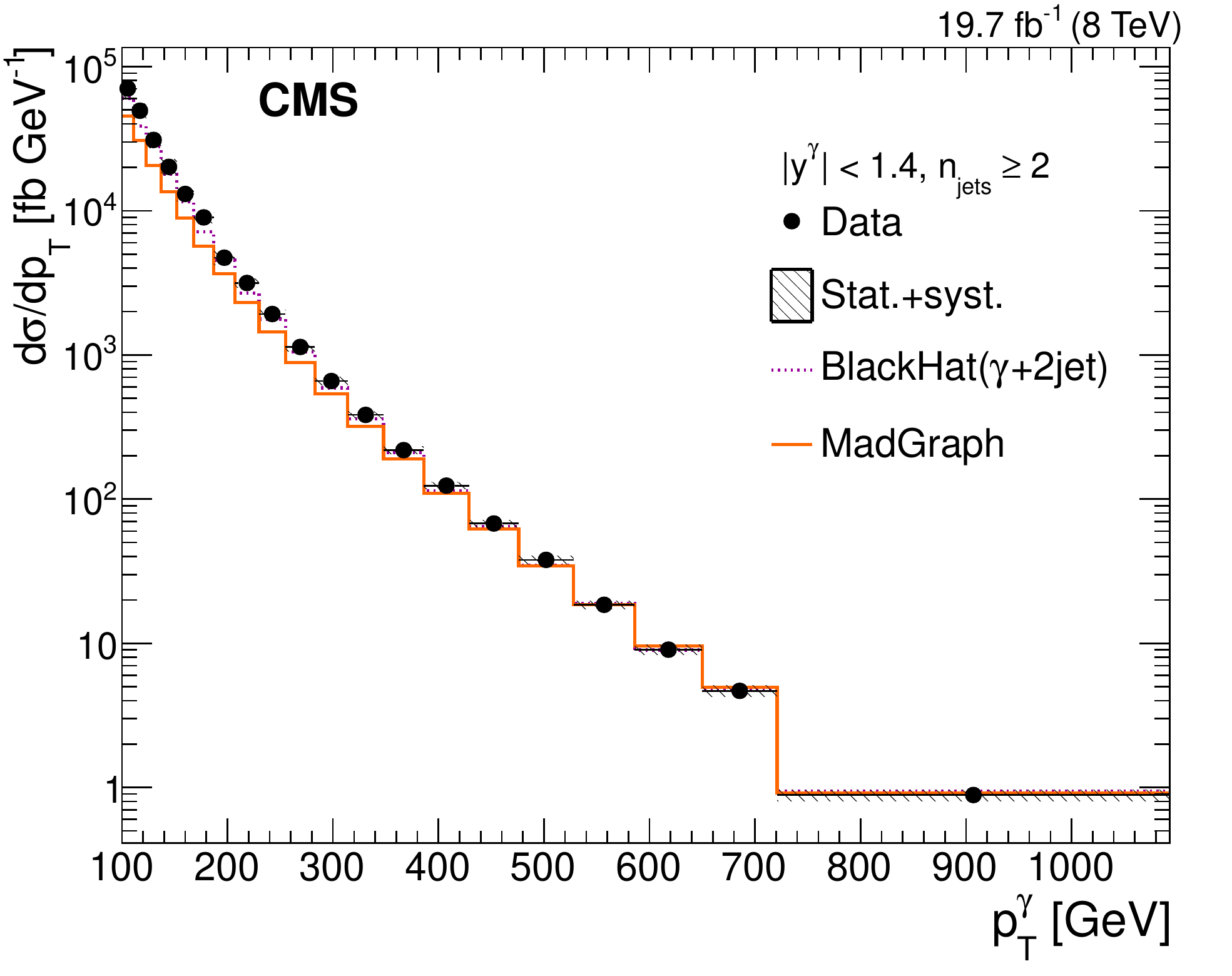}
  \begin{minipage}[c]{0.49\textwidth}
    \includegraphics[width=\textwidth]{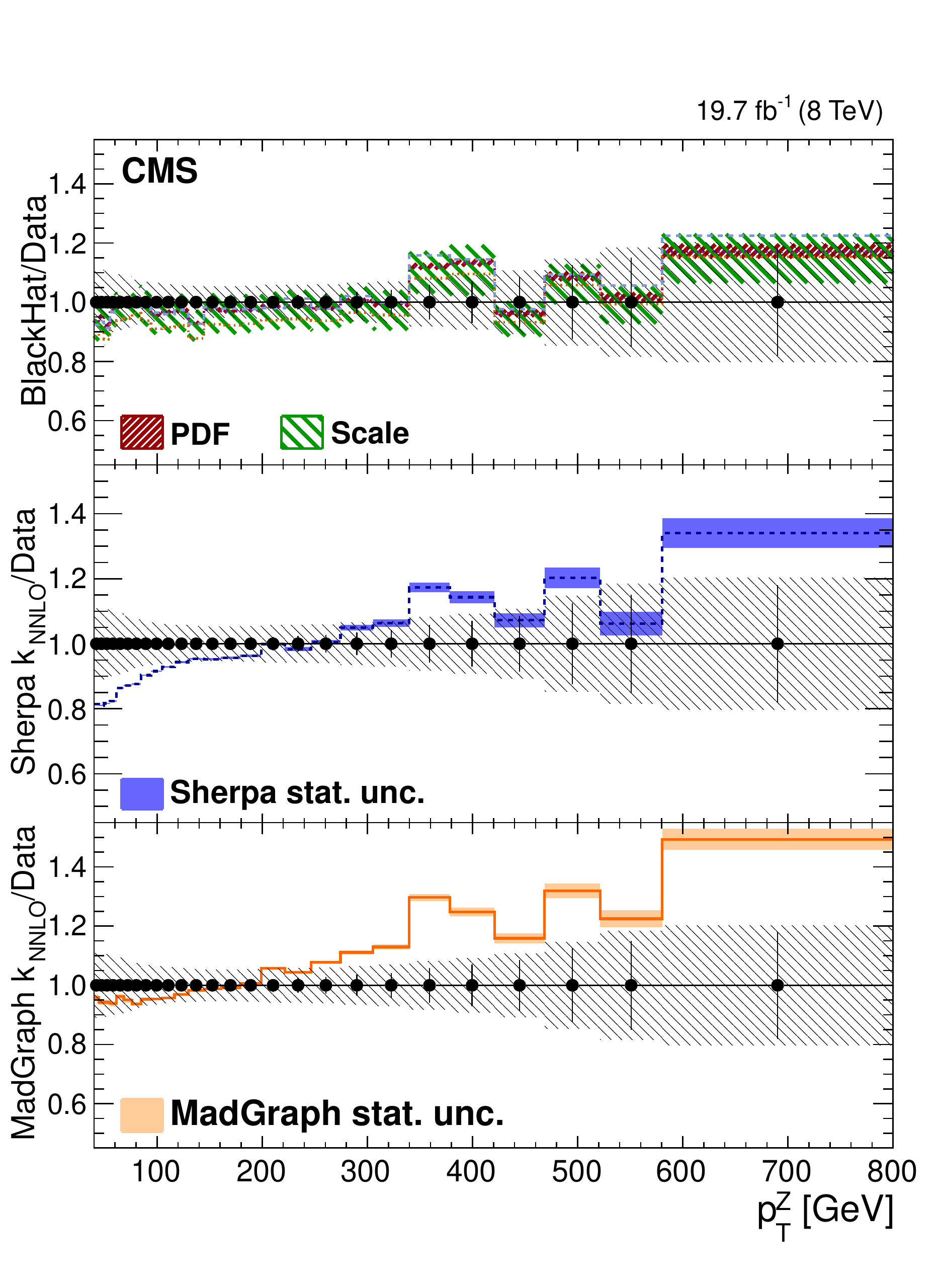}
  \end{minipage}
  \begin{minipage}[c]{0.49\textwidth}
    \centering
    \includegraphics[width=\textwidth]{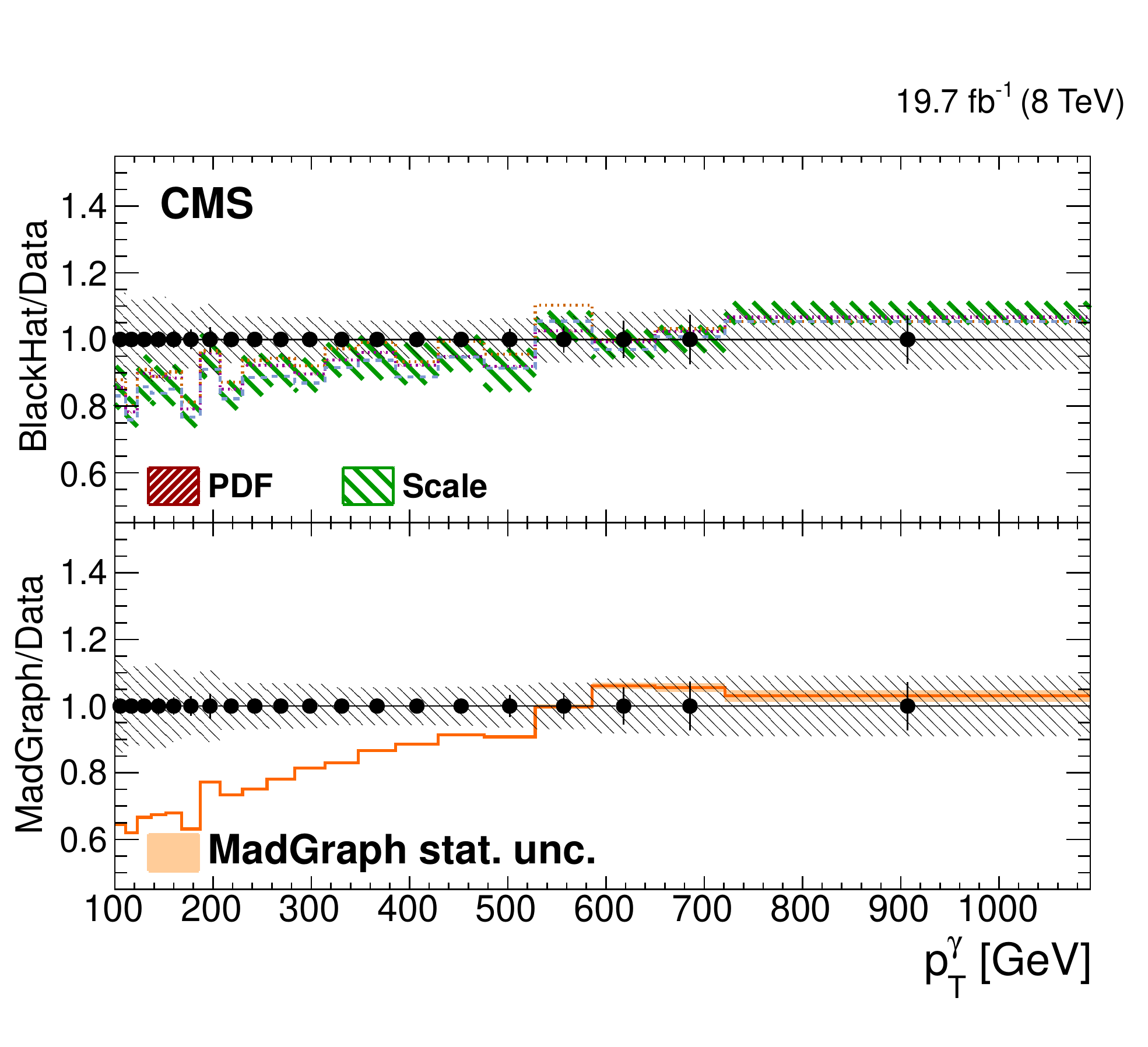}
  \end{minipage}
  \caption{Top left: Differential cross section for \cPZ~boson production as a function of $\ptZ$ for an inclusive \zjets, $\njets\geq2$ selection of detector-corrected data in comparison with estimations from \MADGRAPHPYTHIASIX, \SHERPA, and \BLACKHAT.
Top right: Differential cross section for photon production as a function of $\pt^{\gamma}$ for an inclusive $\gjets$, \mbox{$\njets\geq2$} selection for central rapidities $\abs{y^{\gamma}}<1.4$ in detector-corrected data is compared with estimations from \MADGRAPHPYTHIASIX and \BLACKHAT. A detailed explanation is given in Section~\ref{diffcrosssections}. The bottom plots give the ratio of the various theoretical estimations to the data in the \zjets~case (bottom left) and \gjets~case (bottom right).}
  \label{fig:pT_Bosons_2jet}
\end{figure}

\begin{figure}[hbtp]
  \centering
    \includegraphics[width=0.49\textwidth]{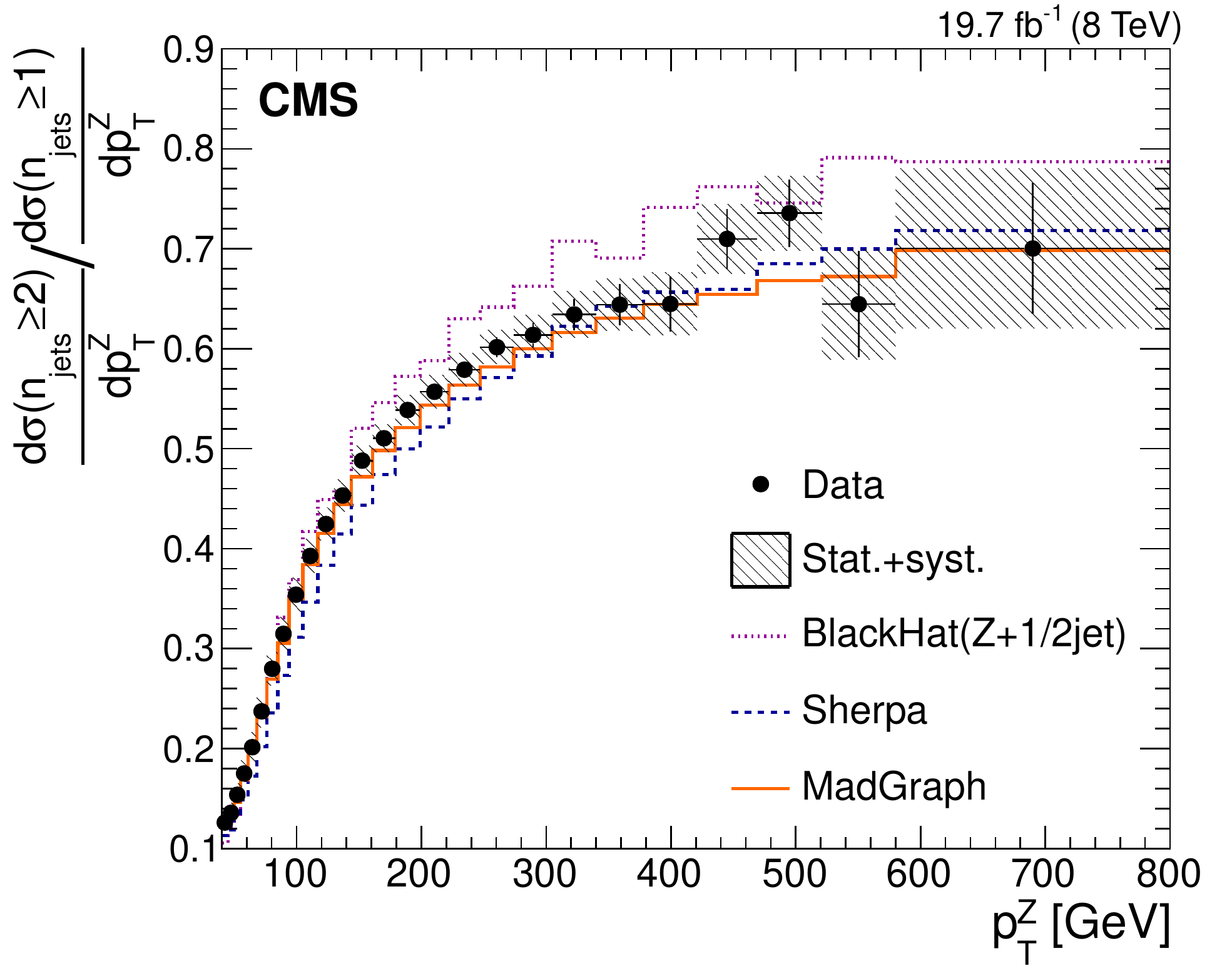}
    \includegraphics[width=0.49\textwidth]{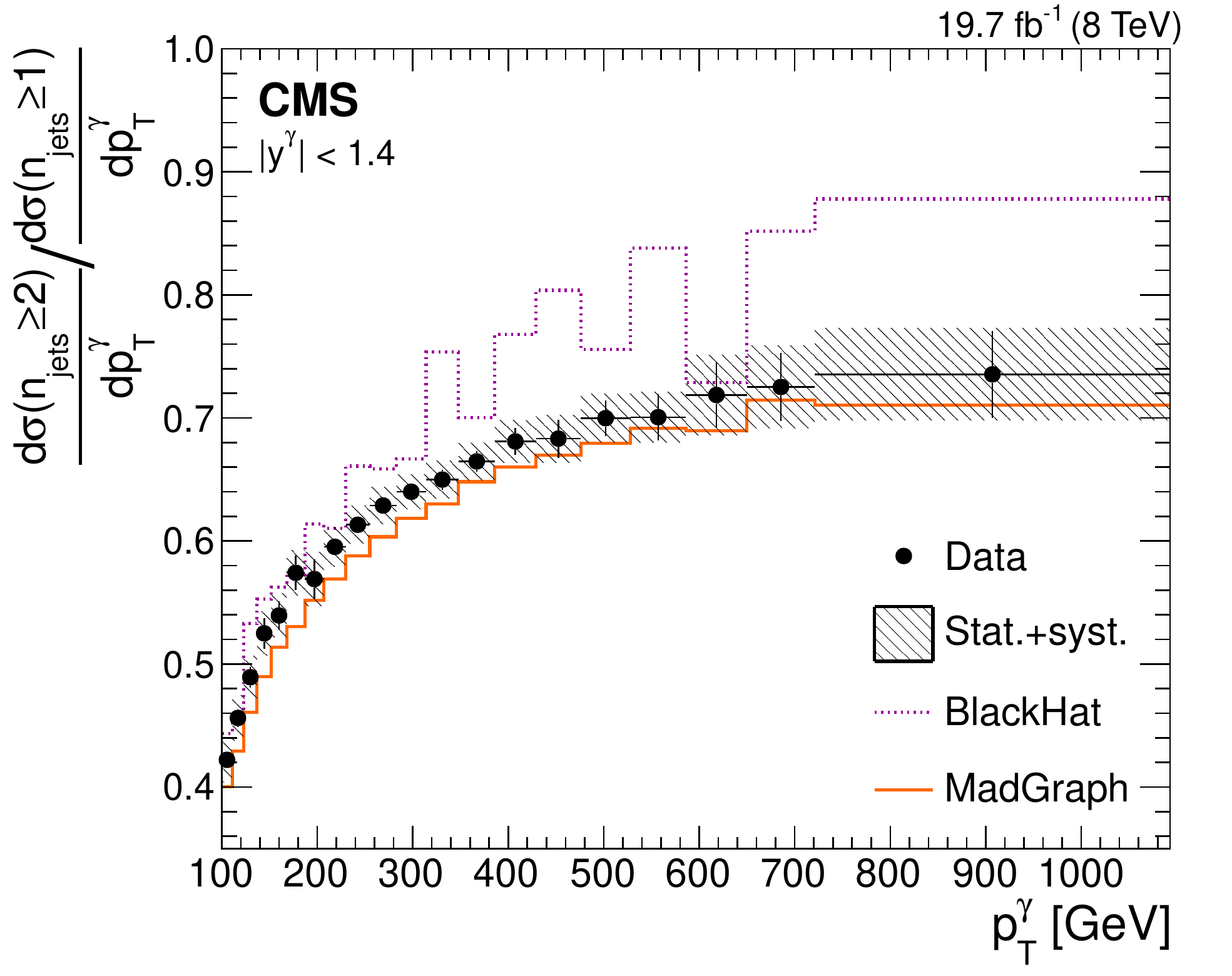}
  \begin{minipage}[c]{0.49\textwidth}
    \centering
    \includegraphics[width=1.0\textwidth]{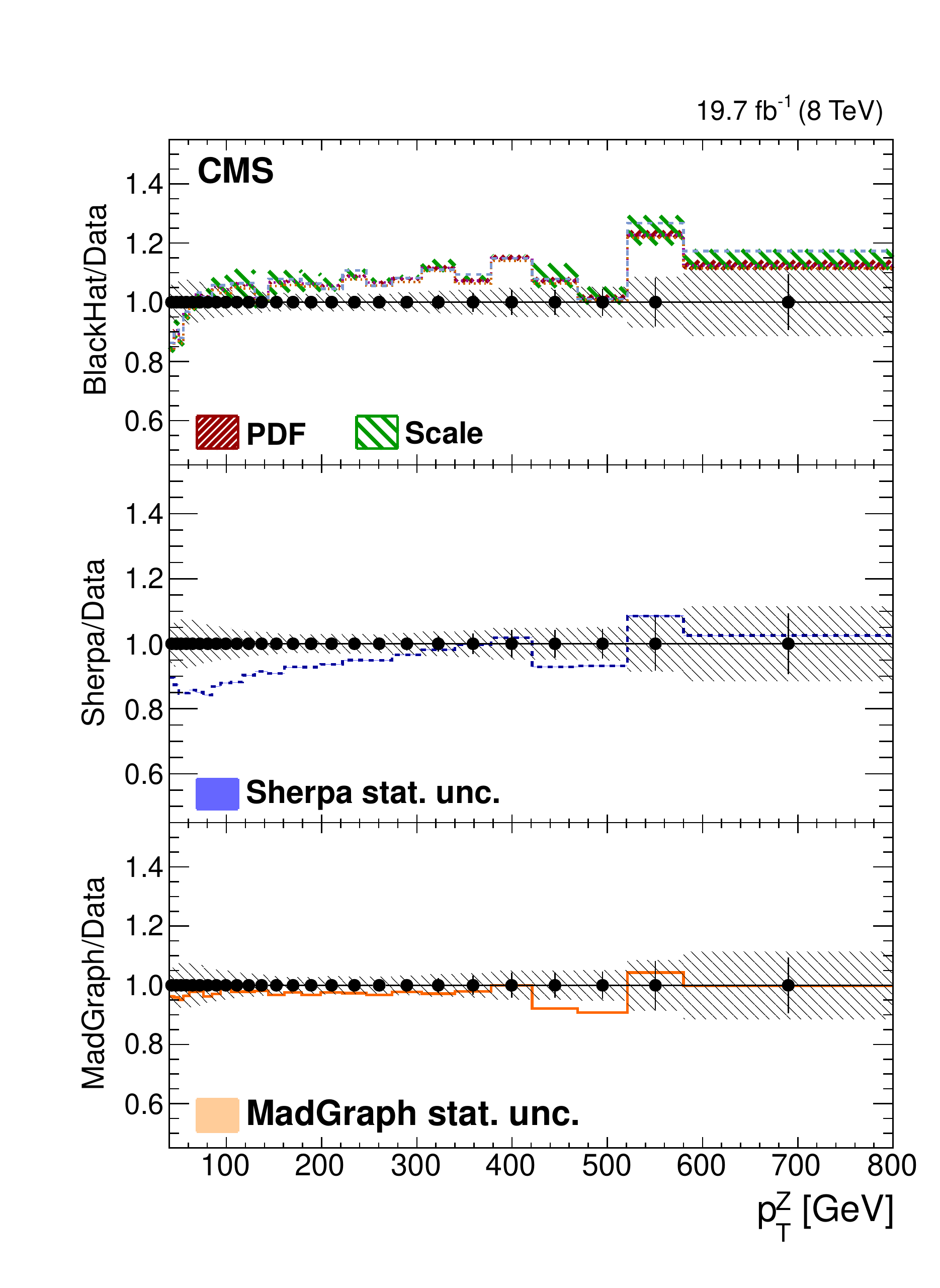}
  \end{minipage}%
  \begin{minipage}[c]{0.49\textwidth}
    \centering
    \includegraphics[width=1.0\textwidth]{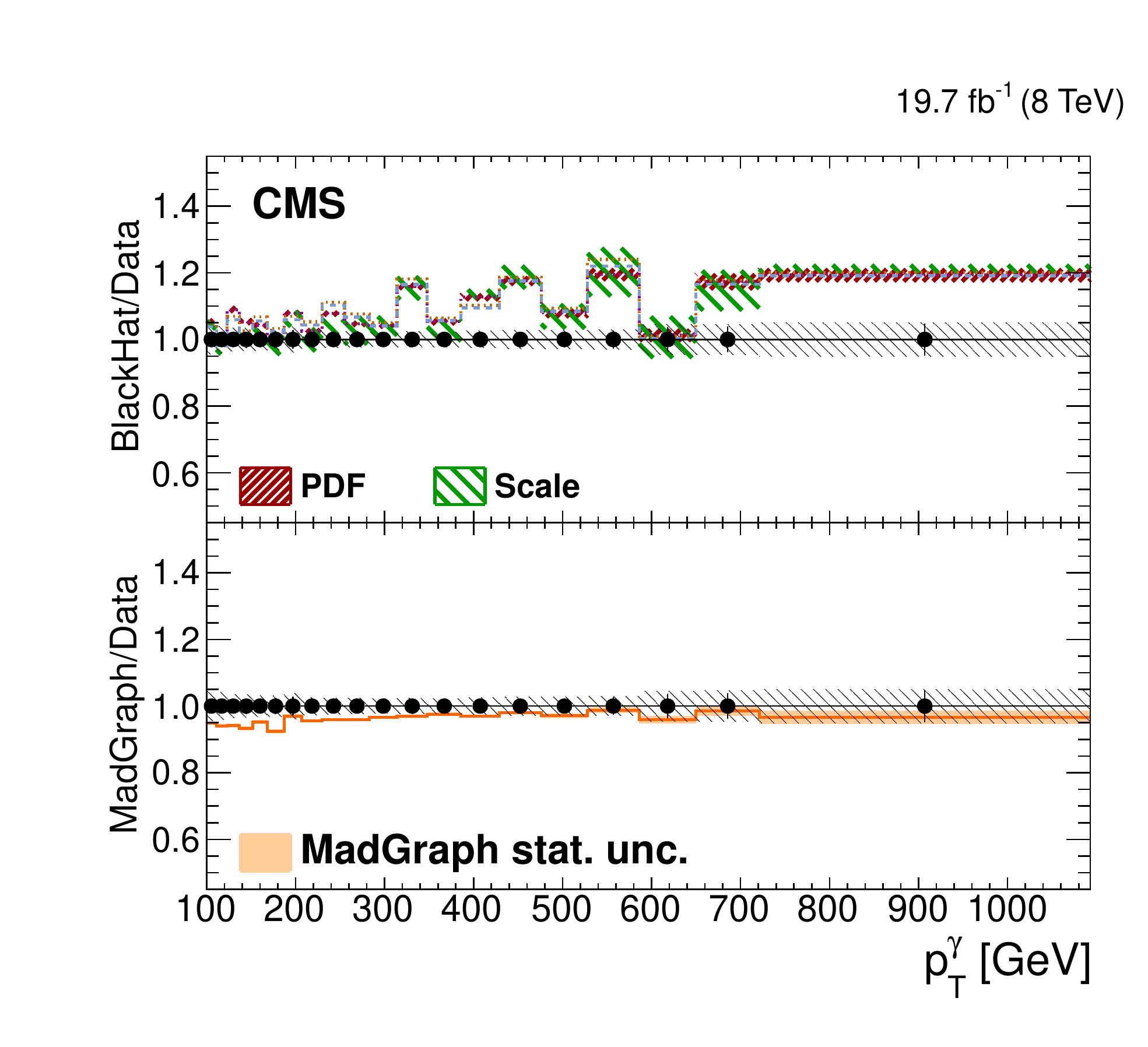}
  \end{minipage}
  \caption{Ratio of the inclusive rates for $\njets\geq2$ and $\njets\geq1$ versus the transverse momentum of the boson for \zjets~in detector-corrected data compared to estimations from \MADGRAPHPYTHIASIX, \SHERPA, and \BLACKHAT (top left) and for $\gjets$ for central rapidities $\abs{y^{\gamma}}<1.4$ in detector-corrected data compared with estimations from \MADGRAPHPYTHIASIX and \BLACKHAT (top right). A detailed explanation is given in Section~\ref{diffcrosssections}. The bottom plots give the ratio of the various theoretical estimations to the data in the \zjets~case (bottom left) and \gjets~case (bottom right).}
  \label{fig:Njetsratios_Bosons}
\end{figure}

In Figs. \ref{fig:pt_Bosons} and \ref{fig:pT_Bosons_2jet}, we present the measured differential cross sections as functions of the $\ptZ$ and the $\pt^{\gamma}$ for two selections of \zjets~and \gjets~events ($\njets\geq1$ and $\njets\geq2$) and compare them with estimates from \BLACKHAT and \MADGRAPHPYTHIASIX. In Fig. \ref{fig:Njetsratios_Bosons}, we present the ratio of the inclusive 2-jet events to the inclusive 1-jet events. For \zjets, we also compare the data to \SHERPA results. The NLO \BLACKHAT estimate is corrected for nonperturbative effects (hadronization and underlying event) using \MADGRAPHPYTHIASIX. These corrections are typically around 2\%.  We use the $n$-jets \BLACKHAT sample for comparison with data and other MC generators in the corresponding inclusive $n$-jets selection. The $\zjets$ simulations from \SHERPA and \MADGRAPHPYTHIASIX are rescaled by a constant NNLO $K$-factor of $K=1.197$, as calculated with \FEWZ 3.1~\cite{FEWZ}, while for \gjets~the LO cross section from \MADGRAPHPYTHIASIX is used as no NNLO $K$-factor is available for \gjets. In all figures, the hatched band surrounding the data points represents the total uncertainty in the measurement, while the error bars show the statistical uncertainty. Similarly, in the MC/data ratio plots, the error bars around the points centered at one represent the relative statistical uncertainties on the data, while the hatched band represents the relative total uncertainty of statistics and systematics on the data. The shaded bands around the MC simulation/data ratios for \MADGRAPHPYTHIASIX and \SHERPA represent the statistical uncertainty (stat. unc.) in the simulation. The outer hatched band around the \BLACKHAT/data ratio (using MSTW2008) shows the total uncertainty of the estimate due to PDF and scale variations, while the inner hatched band indicates the uncertainty due to the variations within the MSTW2008 eigenvector set~\cite{PhysRevD.65.014013}. Analogous variations using the CT10 and NNPDF2.3 PDF sets lead to similar uncertainties. Not shown in the figures is the statistical uncertainty for the \BLACKHAT calculations that amounts to less than 1--3\% for $\njets\geq1,\,2$ and to 5--10\% for $\njets\geq3$ in the $\pt^{Z}$ spectra. In the distributions of the observables $\ptZ/\HT$ and $\log_{10}\left(\ptZ/\pt^{\mathrm{j}1}\right)$, the statistical uncertainty is 6\% except in the tails where there are fewer events. In the $\pt^{\gamma}$ spectra, the statistical uncertainty is 3--5\% in the $\njets\geq1$ case and 4--10\% in the $\njets\geq2$ and 3 cases.  The fluctuations seen in the \BLACKHAT distributions between adjacent bins are statistical in nature. Overlaid are \BLACKHAT estimates using the NNPDF (dashed) and CT10 (dotted) PDF sets.

In the $\zjets$ distributions for both phase space selections ($\njets\geq1$ and $\njets\geq2$, Figs.~\ref{fig:pt_Bosons} and~\ref{fig:pT_Bosons_2jet}), we observe the same qualitative behavior of the ratio of the \MADGRAPHPYTHIASIX simulation to data, which is flat about unity up to around 150--200\GeV and then increases to about 1.3 at higher $\pt$.
Estimates from \SHERPA are lower than the data for $\pt^{\cPZ}<50\GeV$, while for higher $\ptZ$ they increase to around 20\% higher. In the $\njets\geq1$ case, \BLACKHAT shows a flat ratio with respect to data starting around $\pt^{\cPZ}\approx100\GeV$, but underestimates the yield seen in data by 8--10\%, whereas in the $\njets\geq2$ case, \BLACKHAT agrees with the data within the uncertainties for the whole range. For all multiplicity phase space selections, the systematic uncertainty in the MSTW2008 PDF set is 2--3\% in the \BLACKHAT estimate. The central points of CT10 show a difference compared to MSTW2008 of at most 4\%, whereas NNPDF shows a variation of 2\%. The scale uncertainty for MSTW2008 in the \BLACKHAT estimate, as determined through independent variations of the renormalization and factorization scales by factors of 2 and 0.5, leads to an envelope with values of typically 5--10\%.

In the $\njets\geq2$ and $\njets\geq1$ \gjets~case, \BLACKHAT reproduces the shape of the data distribution, but underestimates the rate by approximately 10--15\% throughout most of the range.

In Fig.~\ref{fig:Njetsratios_Bosons}, we see that the inclusive 1-jet over inclusive 2-jet \ptZ~cross section ratio increases until a plateau is reached at around 350\GeV.
The systematic uncertainties are treated as fully correlated in the ratio.
The distributions are well predicted by \MADGRAPHPYTHIASIX in both channels.
\SHERPA underestimates the relative rate of inclusive 2-jets events.
For \BLACKHAT, the inclusive 2-jet generated sample is used to predict the 2-jet rate and to compute the ratio with the predicted rates from the inclusive 1-jet sample; \BLACKHAT overestimates the ratio by  10\% for $\pt^{\cPZ}>100\GeV$ in both the \gjets~and \zjets~cases.

\begin{figure}[hbtp!]
  \centering
\includegraphics[width=0.49\textwidth]{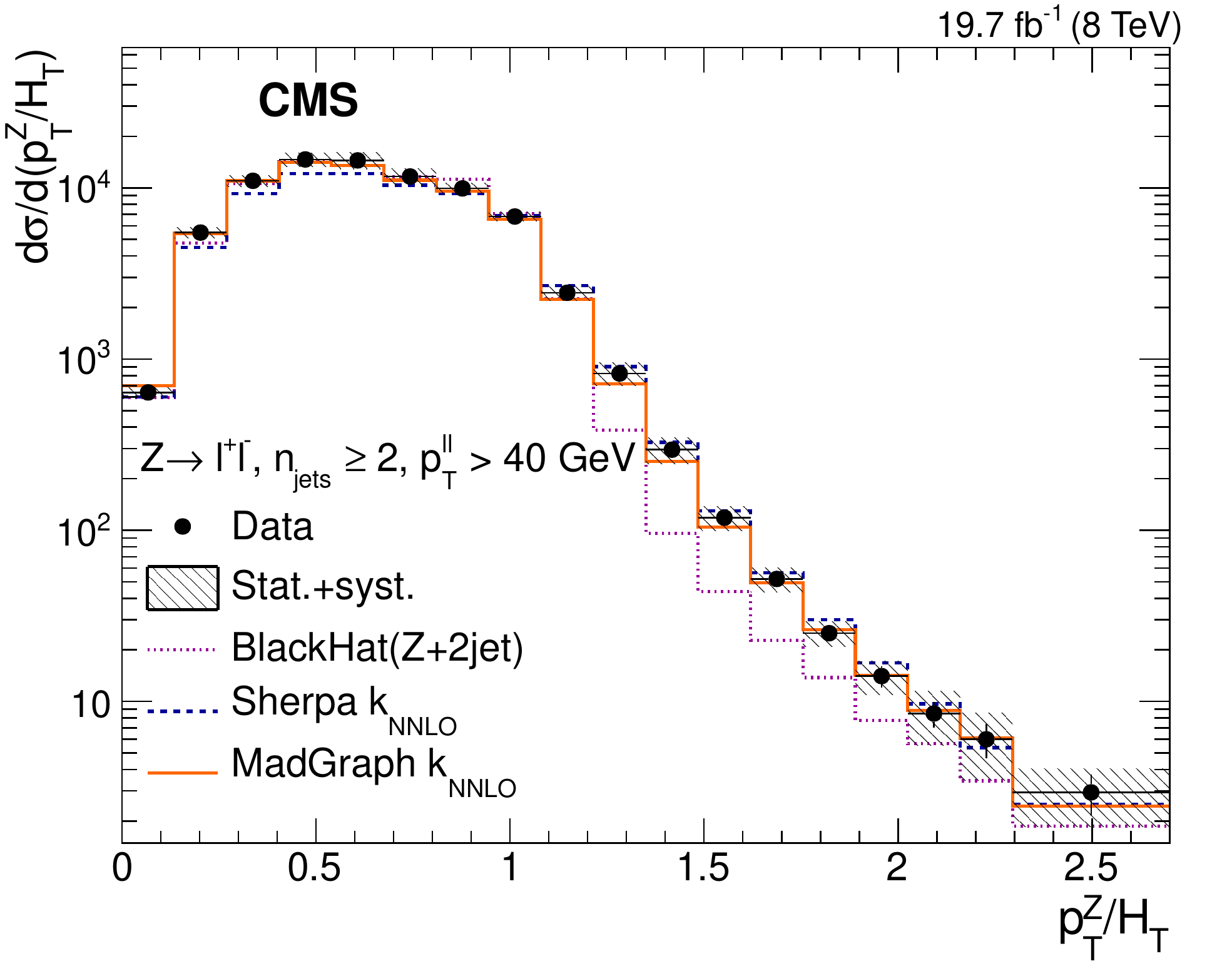}
\includegraphics[width=0.49\textwidth]{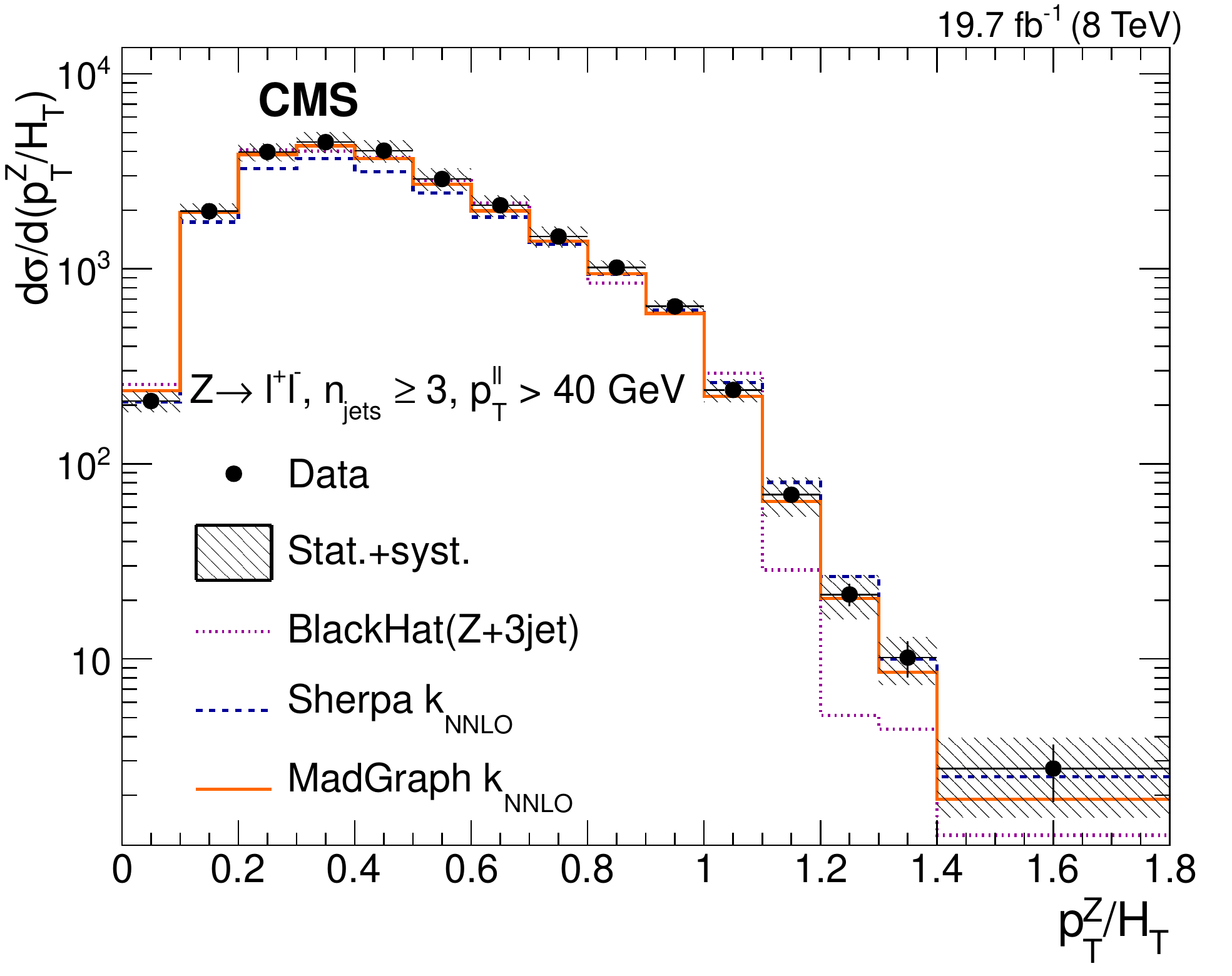}
\begin{minipage}[c]{0.49\textwidth}
 \centering
\includegraphics[width=1.0\textwidth]{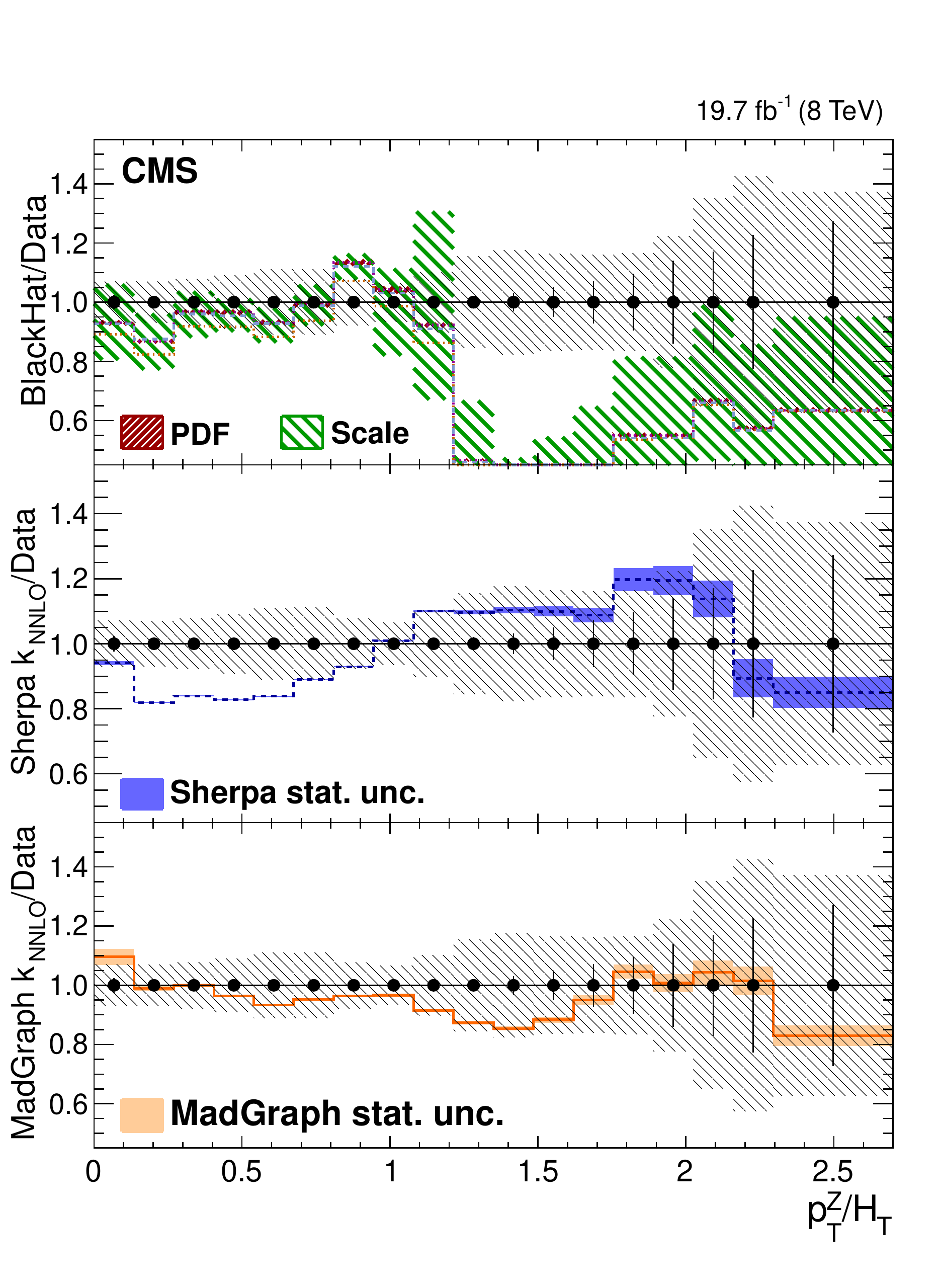}
\end{minipage}%
\begin{minipage}[c]{0.49\textwidth}
 \centering
\includegraphics[width=1.0\textwidth]{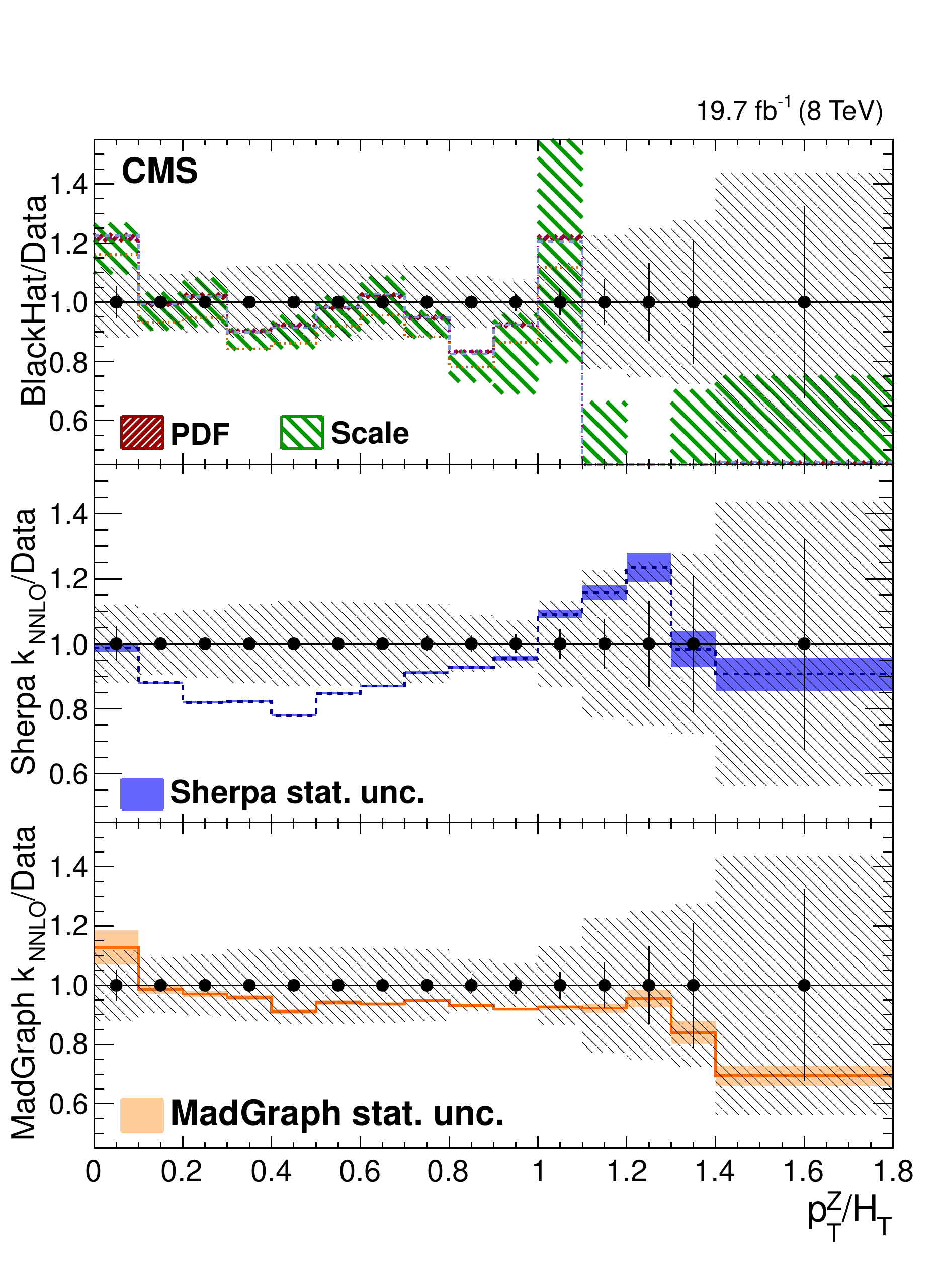}
\end{minipage}
     \caption{The measured distribution of the observable $\pt^{\cPZ}/\HT$ ratio for $\njets\geq2$ (top left) and $\njets\geq3$ (top right) for \zjets~in detector-corrected data compared with estimations from \MADGRAPHPYTHIASIX,
\SHERPA, and \BLACKHAT.  A detailed explanation is given in Section~\ref{diffcrosssections}. The bottom plots give the ratio of the various theoretical estimations to the data in the $\njets\geq2$ case (bottom left) and $\njets\geq3$ case (bottom right).}
         \label{fig:finalmeasurement_ptZ_over_HT}
\end{figure}

\begin{figure}[hbtp]
  \centering
\includegraphics[width=0.49\textwidth]{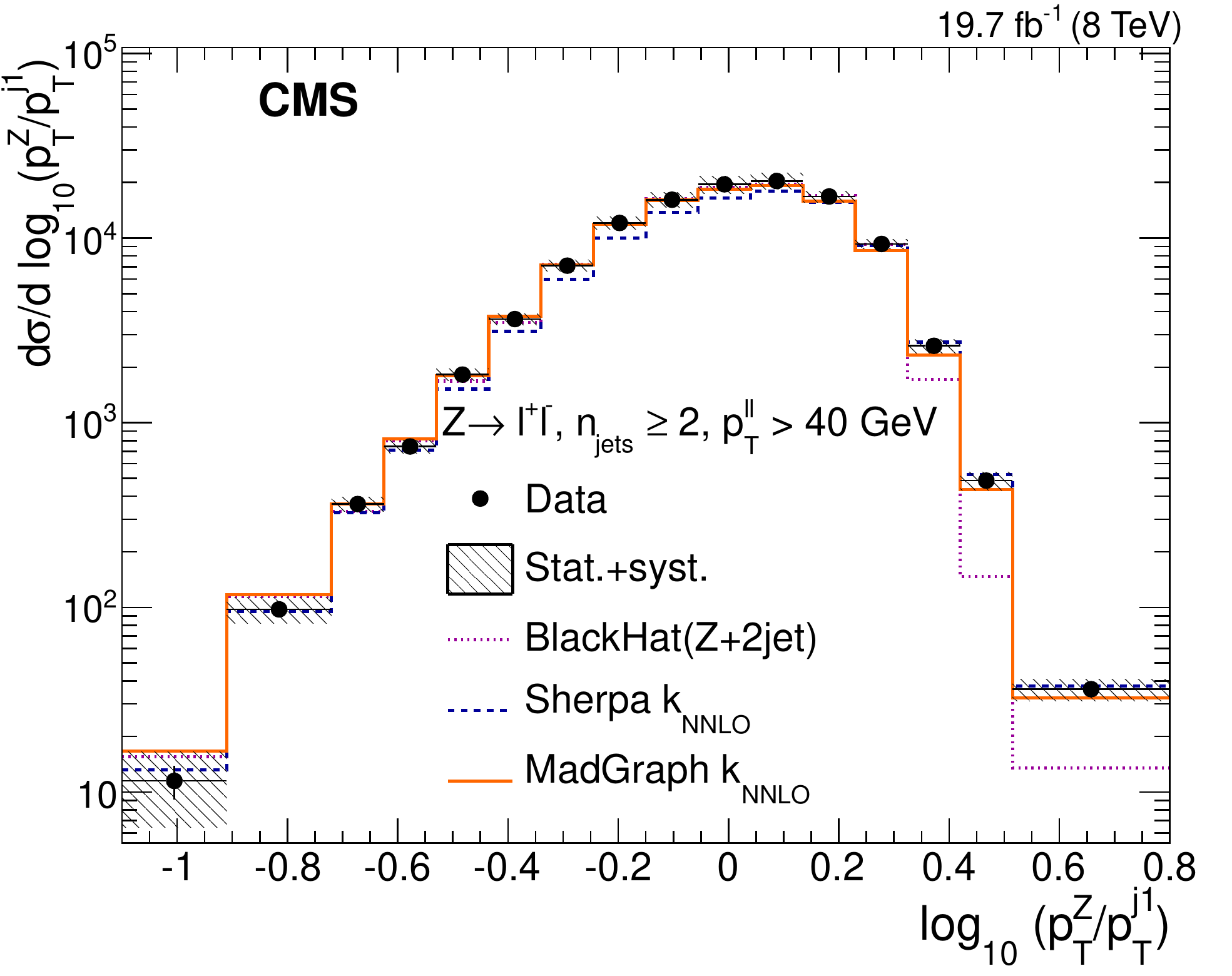}
\includegraphics[width=0.49\textwidth]{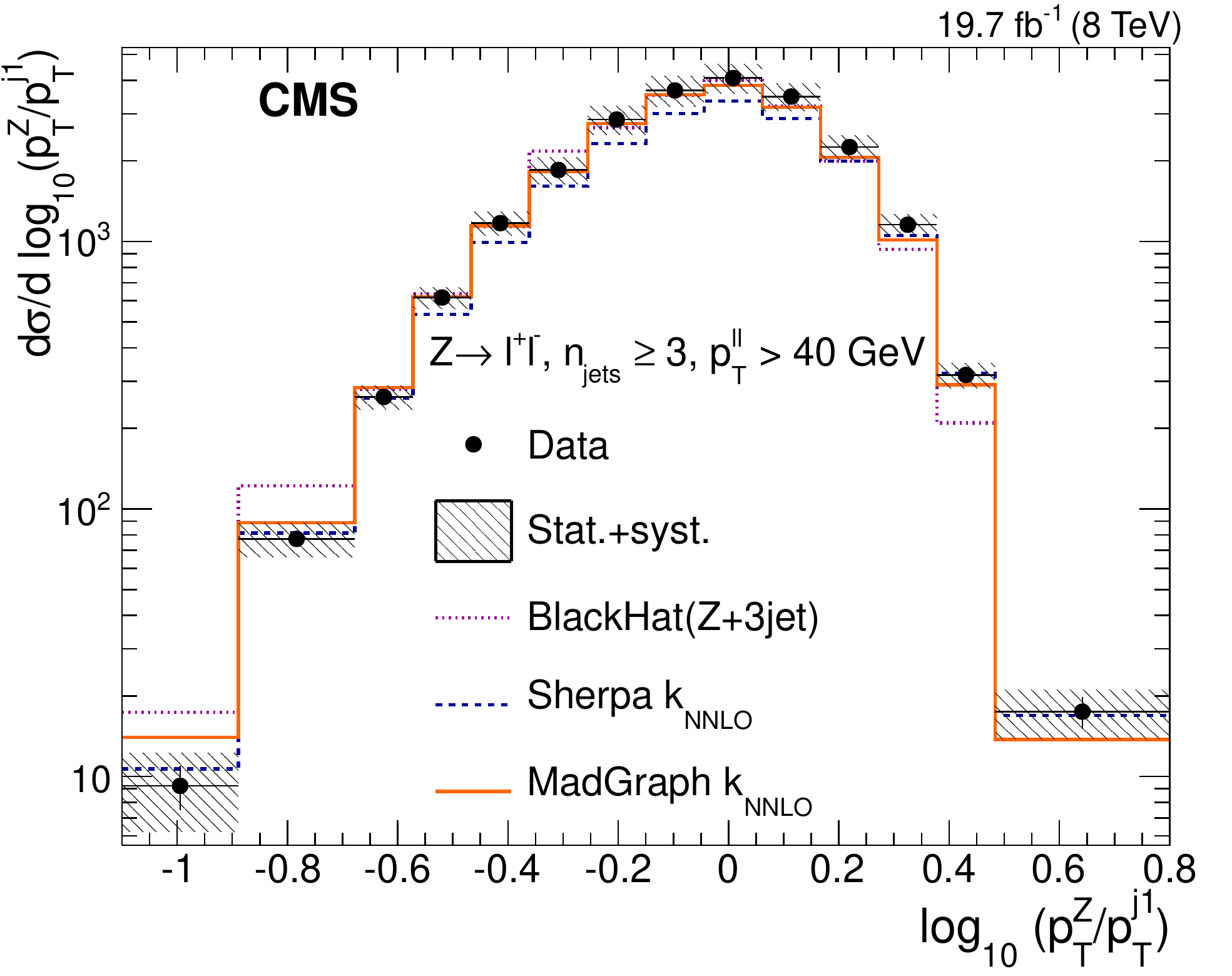}
\includegraphics[width=0.49\textwidth]{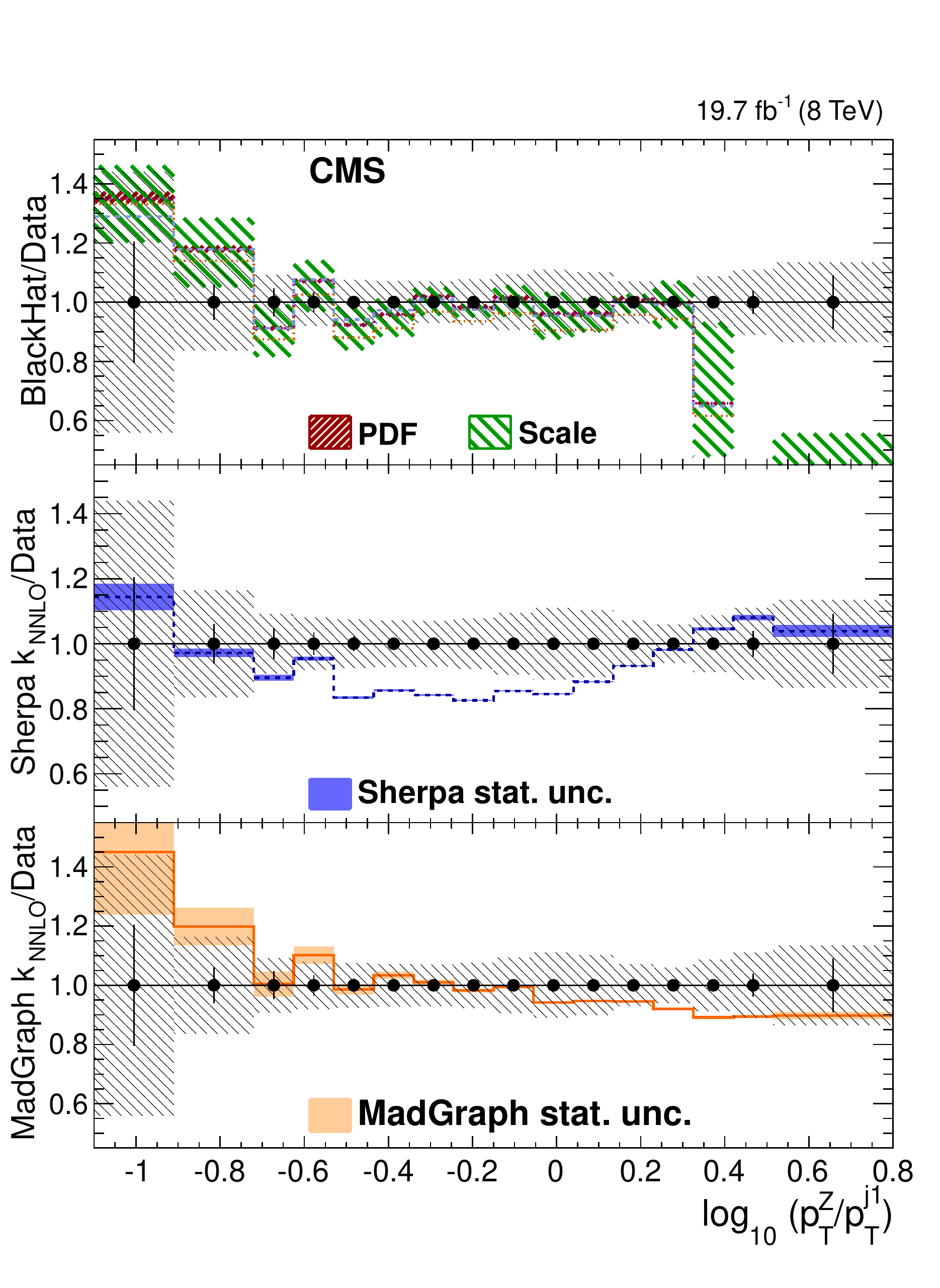}
\includegraphics[width=0.49\textwidth]{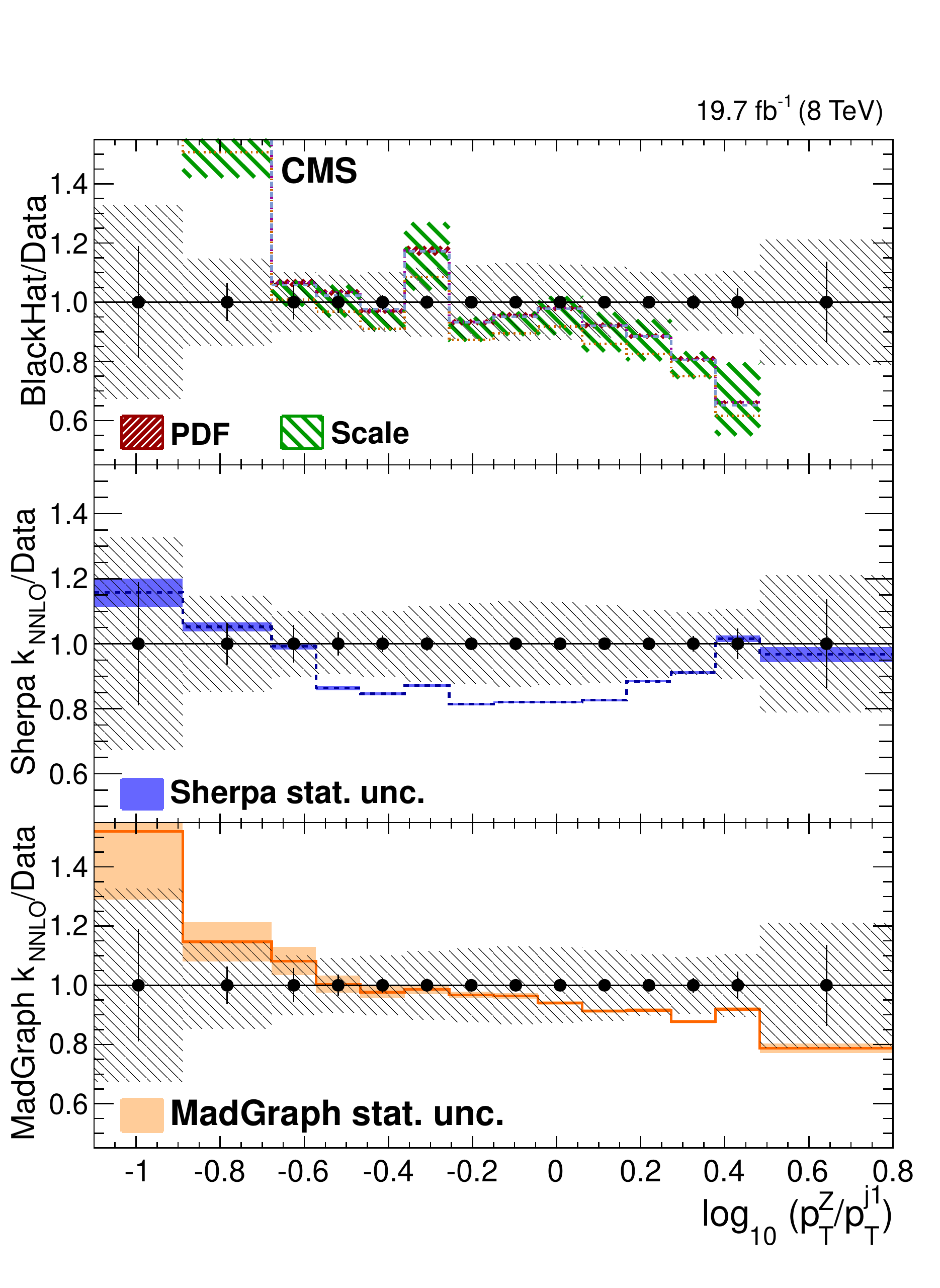}
     \caption{The measured distribution of the observable $\log_{10}\pt^{\cPZ}/\pt^{\mathrm{j}1}$ ratio for $\njets\geq2$ (top left) and $\njets\geq3$ (top right) for \zjets~in detector-corrected data compared with estimations from \MADGRAPHPYTHIASIX, \SHERPA, and \BLACKHAT.  A detailed explanation is given in Section~\ref{diffcrosssections}. The bottom plots give the ratio of the various theoretical estimations to the data in the $\njets\geq2$ case (bottom left) and $\njets\geq3$ case (bottom right).}
         \label{fig:finalmeasurement_log10_ptZ_over_pt1}
\end{figure}

For \zjets, we study the variables $\pt^{\cPZ}/\HT$, shown in Fig.~\ref{fig:finalmeasurement_ptZ_over_HT}, and $\log_{10}\left(\pt^{\cPZ}/\pt^{\mathrm{j}1}\right)$, shown in Fig.~\ref{fig:finalmeasurement_log10_ptZ_over_pt1}, which allow us to test the validity of NLO estimations.  In particular, we examine these distributions as quantities where NLO estimations might reach their calculational limit due to large logarithms or where missing higher-order effects could play a larger role.

For events which contain a dominant high-\pt jet, $\pt^{\cPZ}/\HT$ tends to unity as the jet carries most of the $\pt$ of the event. Events that populate the high-end tail of the distribution have either additional jets outside of
the acceptance in the forward region or additional hadronic radiation that is not clustered in jets with $\pt^{\text{jet}}> 30\GeV$. In hadronic searches for new physics, these events
contribute to signatures with a high $\ETslash/\HT$ ratio. Almost all events with two or more jets
inside the jet acceptance selections have $\ptZ/\HT$ values below one.  This behavior can be observed in
Fig.~\ref{fig:finalmeasurement_ptZ_over_HT}:
increasing the number of required jets leads to a shift of the complete distribution
towards lower values. The nonperturbative corrections are slightly larger, typically below 5\% in the bulk of the distribution, reaching 10\% in the tails for all variables examined here. Overall, \MADGRAPHPYTHIASIX predicts the rate and shape best up to the tails,
while \SHERPA shows differences in both shape and rates. The \BLACKHAT generator performs well
for the bulk of the distribution, but fails to reproduce the tails. This is especially evident in the high-end tail of the distribution, where we see a sharp drop in the ratio of \BLACKHAT simulations to data. In this portion of phase space, \BLACKHAT is effectively reduced from an NLO to LO calculation as the $n$+1 jet LO calculation in the inclusive $n$ jet case dominates here, whereas the other portions provide negligible contributions. This feature is also confirmed by the sharp increase of scale uncertainties in \BLACKHAT estimates, which have a step-like increase from below 10\% to around 60\% at this point. Therefore, this sharp change in the \BLACKHAT over data ratio (e.g., around 1.2 and 1.1 in Fig.~\ref{fig:finalmeasurement_ptZ_over_HT}) is expected and indicates the ``boundary'' between the regions where a fixed-order calculation gives a suitable estimation and where we would need the parton showering to add soft jets or jets in the forward regions of the detector. Additionally, we can use this to check for any large logarithmic contributions in the lower end of this $\pt^{\cPZ}/\HT$ distribution. We see from the agreement in both the 2- and 3-jet cases that there is no evidence of any such contributions.

The distribution of the second variable, $\log_{10} (\pt^{\cPZ}/\pt^{\mathrm{j}1})$, shown in Fig.~\ref{fig:finalmeasurement_log10_ptZ_over_pt1}, shows similar behavior.
For events with exactly one jet, the $\cPZ$ boson and the jet are back-to-back,
with $\pt^{\cPZ}\approx \pt^{\mathrm{j}1}$, and the distribution peaks around zero. Events where the $\cPZ$ boson is the dominating object will have positive values. If the $\cPZ$ boson carries
less $\pt$ than most of the jets, the variable has negative values.
With increasing jet multiplicity the distribution still peaks around zero, but broadens. Figure~\ref{fig:finalmeasurement_log10_ptZ_over_pt1} shows a comparison of MC estimates to data, which is unfolded to particle level.
The \MADGRAPHPYTHIASIX calculation performs well in estimating the behavior of the data for all inclusive multiplicty selections, but there is a slope within uncertainties in the MC/data plot. On the other hand, \BLACKHAT performs well in the middle range, but the behavior in the tails indicates that jet production due to higher-order diagrams is missing.
In the $\zjets$, $\njets \geq2$ phase space we observe a drop
at the value $\log_{10}(\pt^{\cPZ}/\pt^{\mathrm{j}1})=0.3\approx\log_{10}(2)$, corresponding to
$\pt^{\mathrm{j}1}\approx\pt^{\mathrm{j}2}$ with both jets recoiling against the $\cPZ$ boson direction. The distribution drops at the point where the third-leading jet
becomes relevant. Since we use the inclusive 2-jet \BLACKHAT sample in that phase space, 3-jet events are only available as LO contributions in the real part. Therefore, the estimation is effectively an LO calculation at that point onwards, and subsequently becomes less precise and the scale variation uncertainty increases to around 30\% at that point.

\subsection{ \texorpdfstring{The $\cPZ/\gamma$ ratio}{The Z/g ratio}}
\label{resultZGamma}

In order to compare the cross sections for $\zjets$ and $\gjets$,
the rapidity range of the bosons
is restricted to $\abs{y^{V}}<1.4$ because this is the selected
kinematic region for the photons.
The ratio of the differential cross sections as a function of $\pt$ is measured
in the four phase space regions:
$\njets\geq 1,\,2,\,3$, and $\HT>300\GeV$, $\njets\geq1$.

Statistical uncertainties in the ratio are propagated using the diagonal terms of the covariance matrices. The sources of systematic uncertainty such as the JES, the luminosity uncertainty, and the JER are correlated between \zjets~and \gjets~and therefore cancel in the ratio.
The remaining uncertainties are results of the photon purity measurement, unfolding uncertainty, the uncertainties in the efficiency determination for photons, and the lepton energy or momentum scale uncertainty.

\begin{figure}[hbpt]
  \centering
  \includegraphics[width=0.49\textwidth]{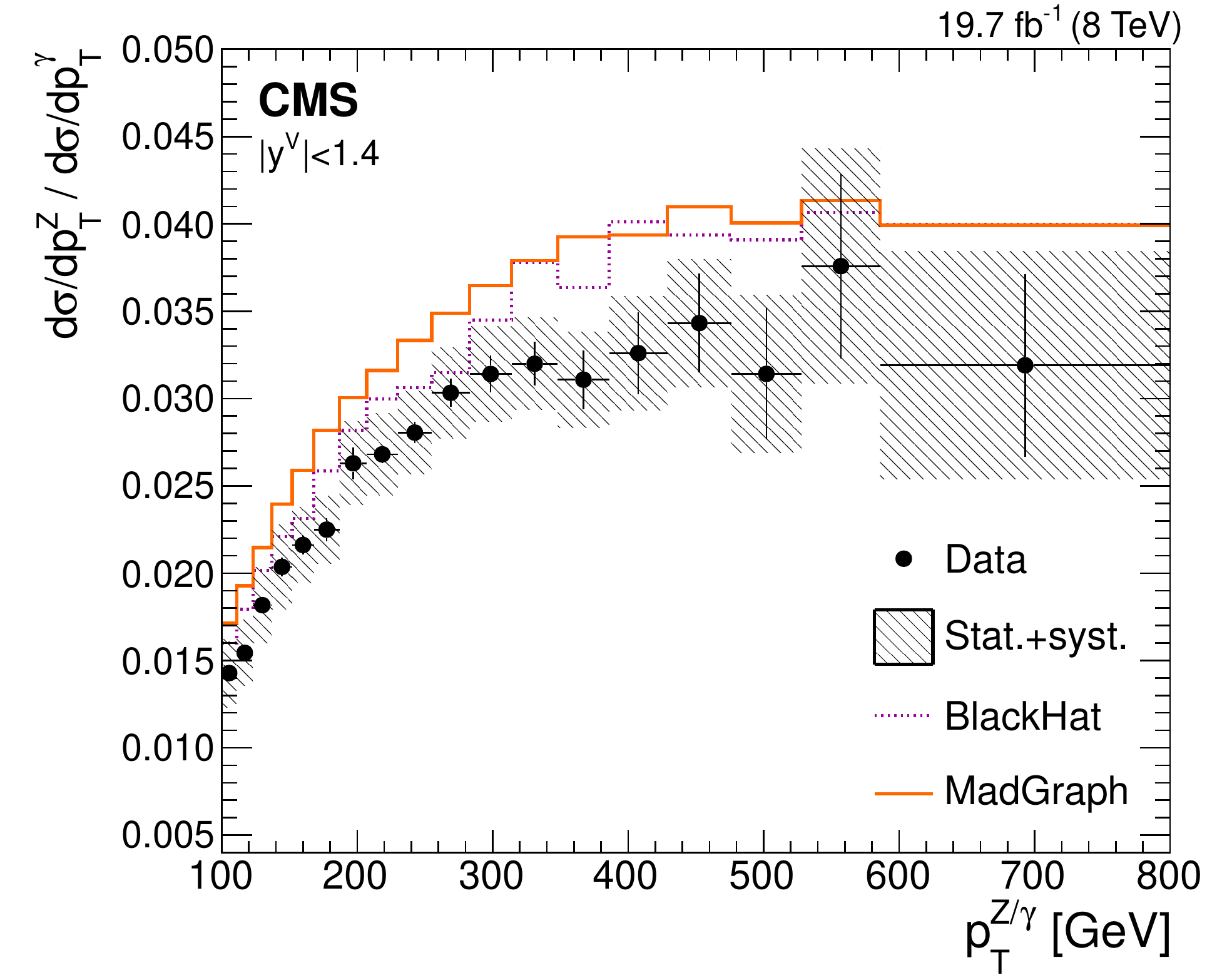}
  \includegraphics[width=0.49\textwidth]{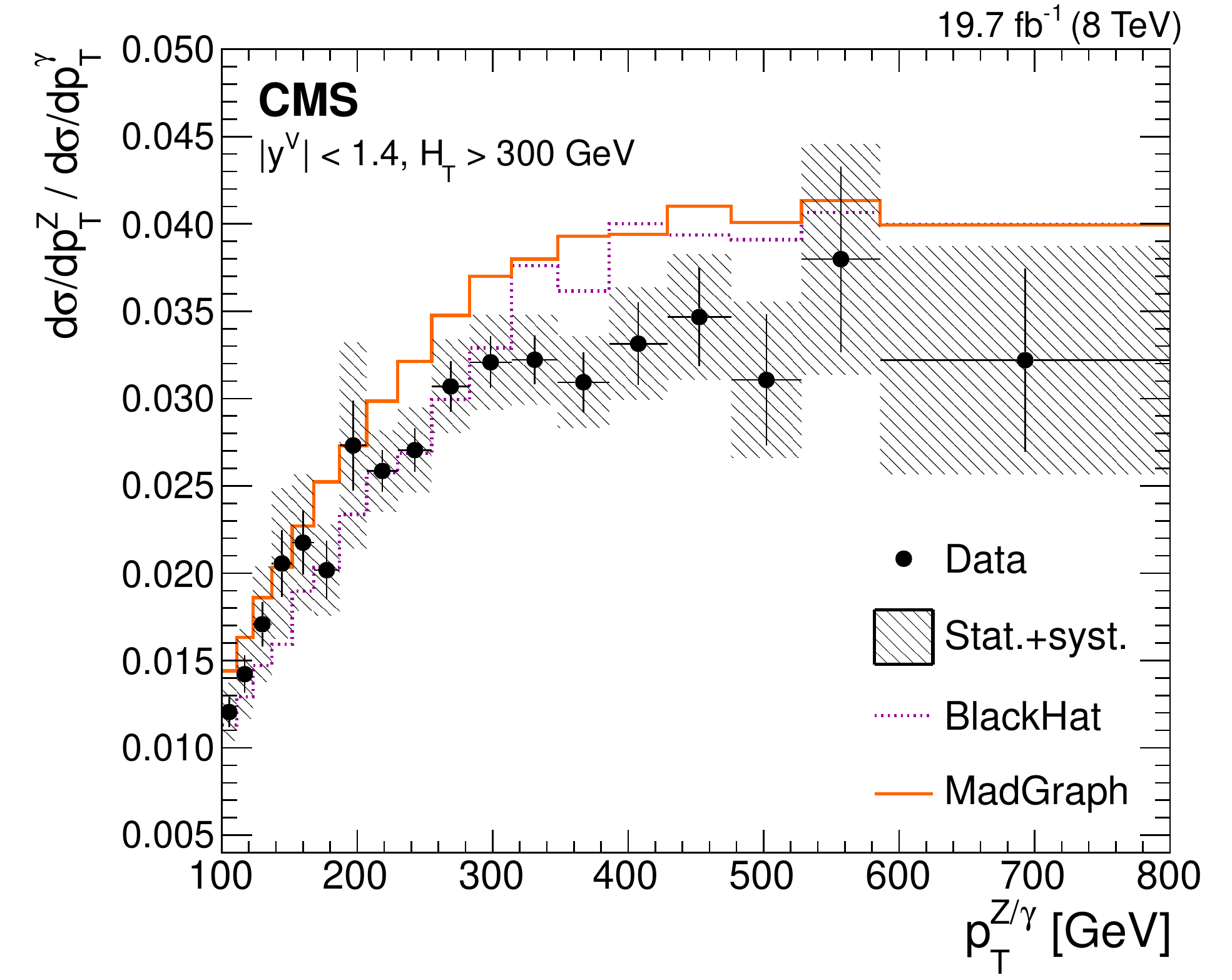}
  \includegraphics[width=0.49\textwidth]{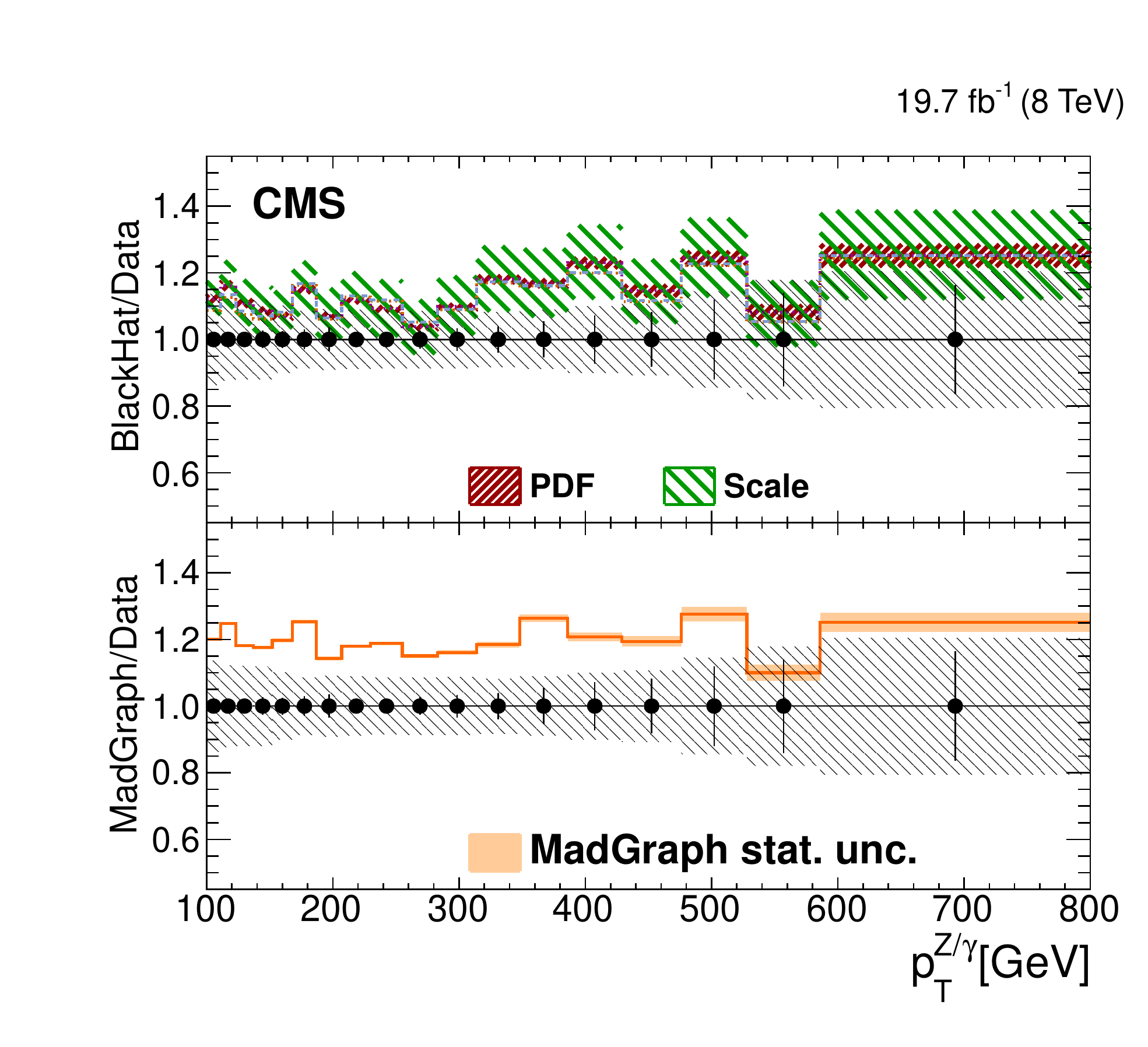}
  \includegraphics[width=0.49\textwidth]{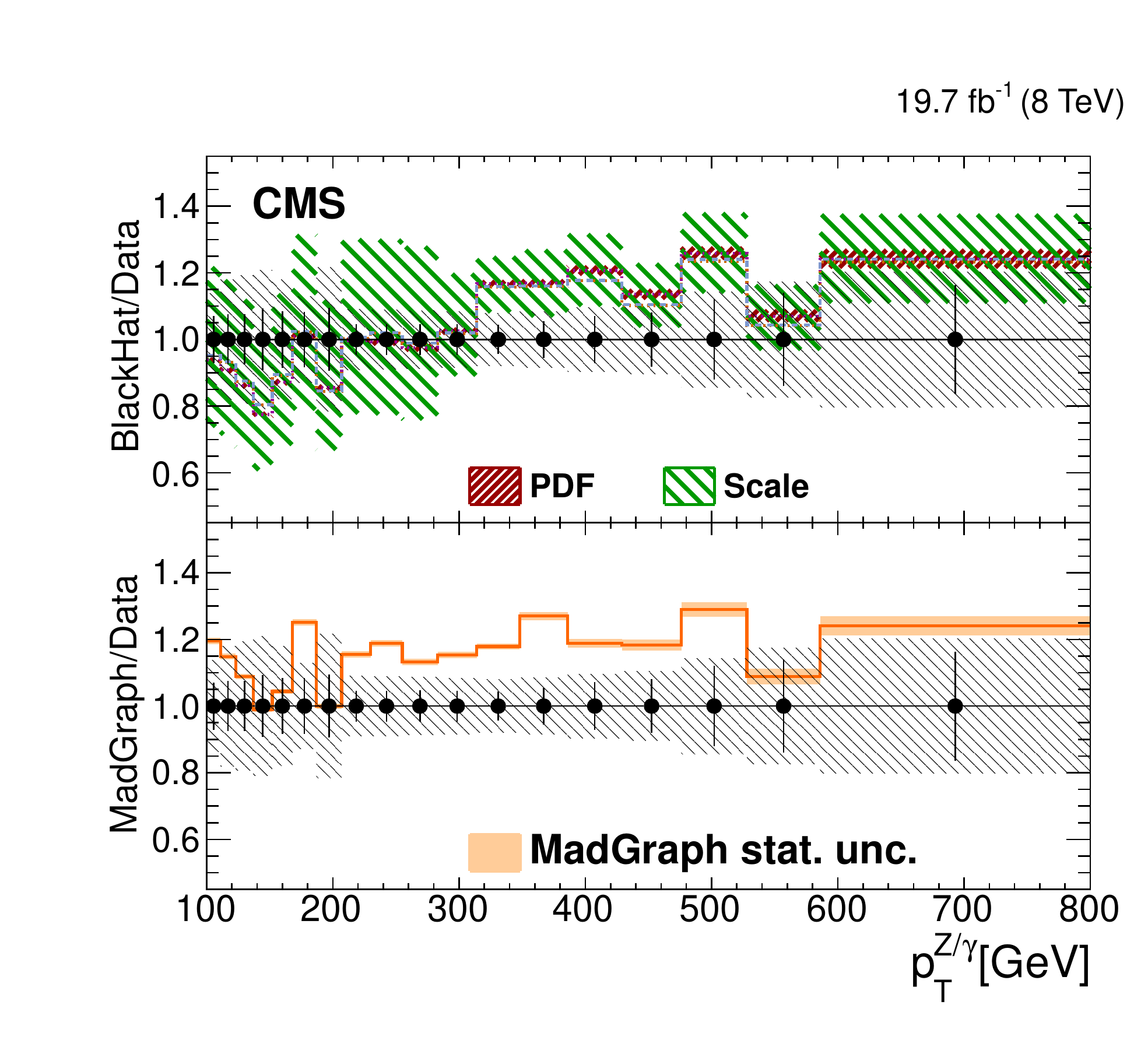}
  \caption{Differential cross section ratio of averaged $\cPZ\to\left(\Pep\Pem+\PGmp\PGmm\right)$ over $\gamma$ as a function
of the total transverse-momentum cross section and for central bosons ($\abs{y^{V}}<1.4$) at different kinematic selections in detector-corrected data. Top left: inclusive ($\njets\ge1$); top right: $\HT\ge300\GeV$, $\njets\ge1$. The black error bars reflect the statistical uncertainty in the ratio, the hatched (gray) band represents the total uncertainty in the measurement. The shaded band around the \MADGRAPHPYTHIASIX simulation to data ratio represents the statistical uncertainty in the MC estimation. The bottom plots give the ratio of the various theoretical estimations to the data in the $\njets\geq1$ case (bottom left) and $\HT\geq300\GeV$ case (bottom right).}
   \label{fig:Zpt_over_gamma_pt}
\end{figure}

\begin{figure}[hbpt]
  \centering

  \includegraphics[width=0.49\textwidth]{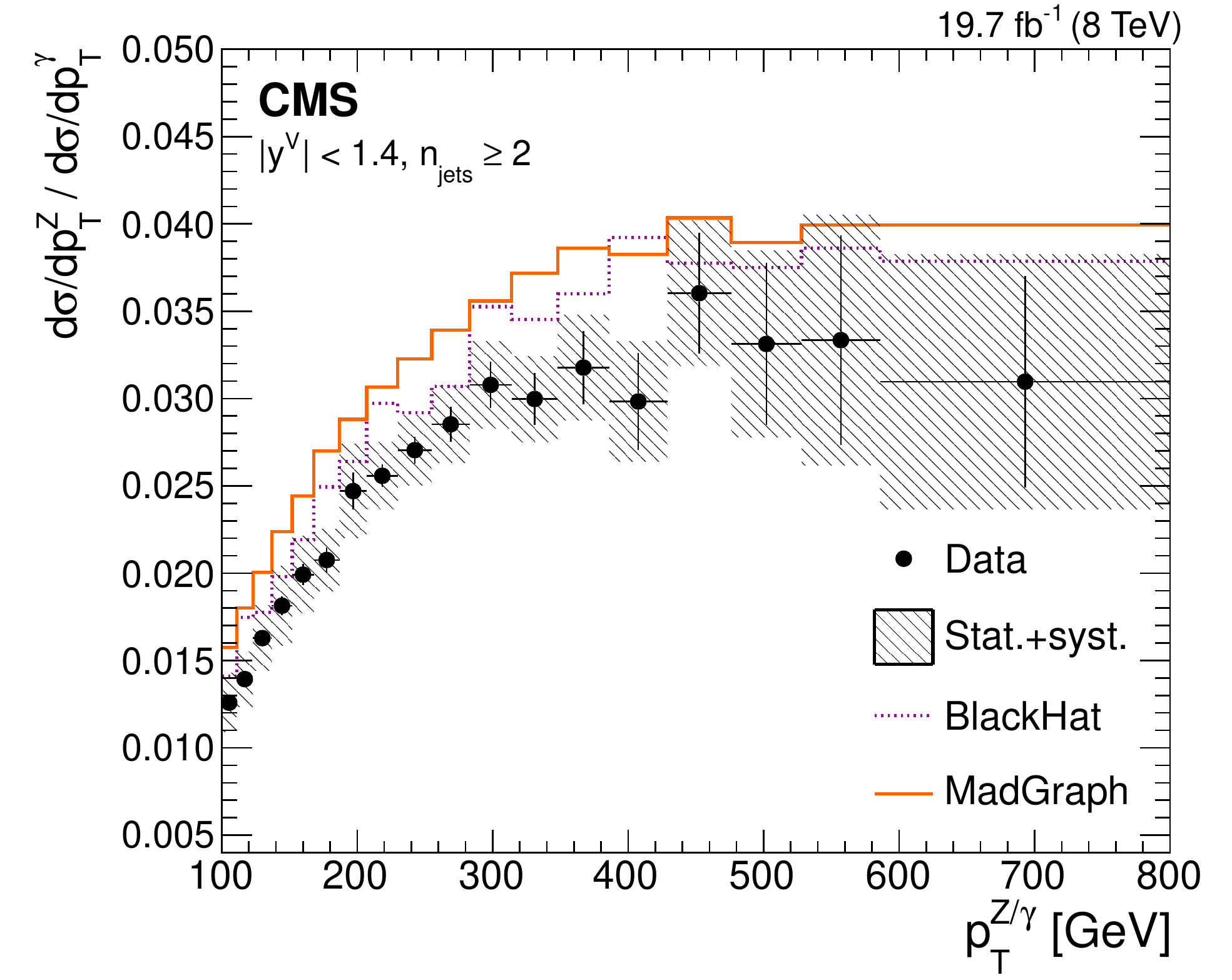}
  \includegraphics[width=0.49\textwidth]{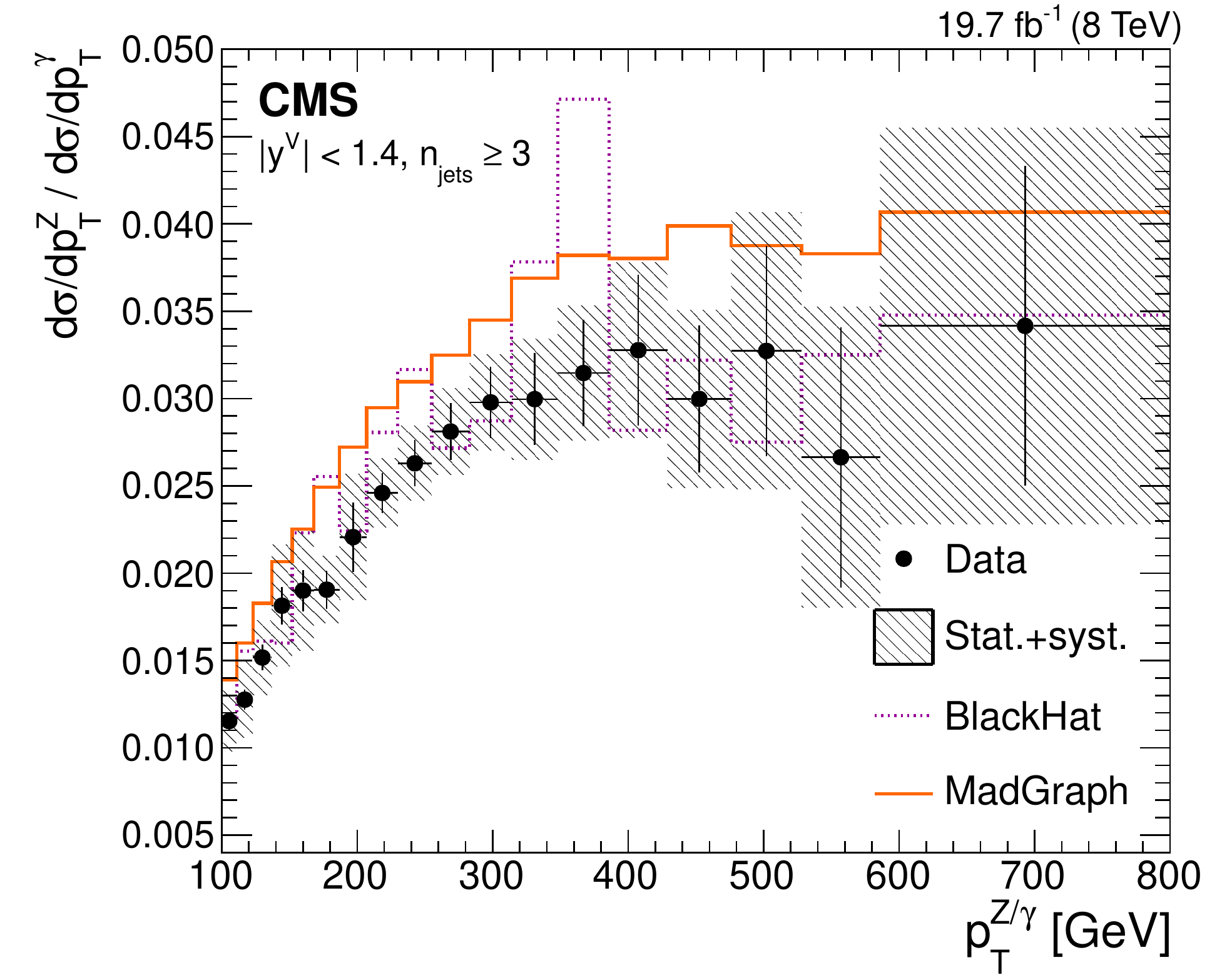}
  \includegraphics[width=0.49\textwidth]{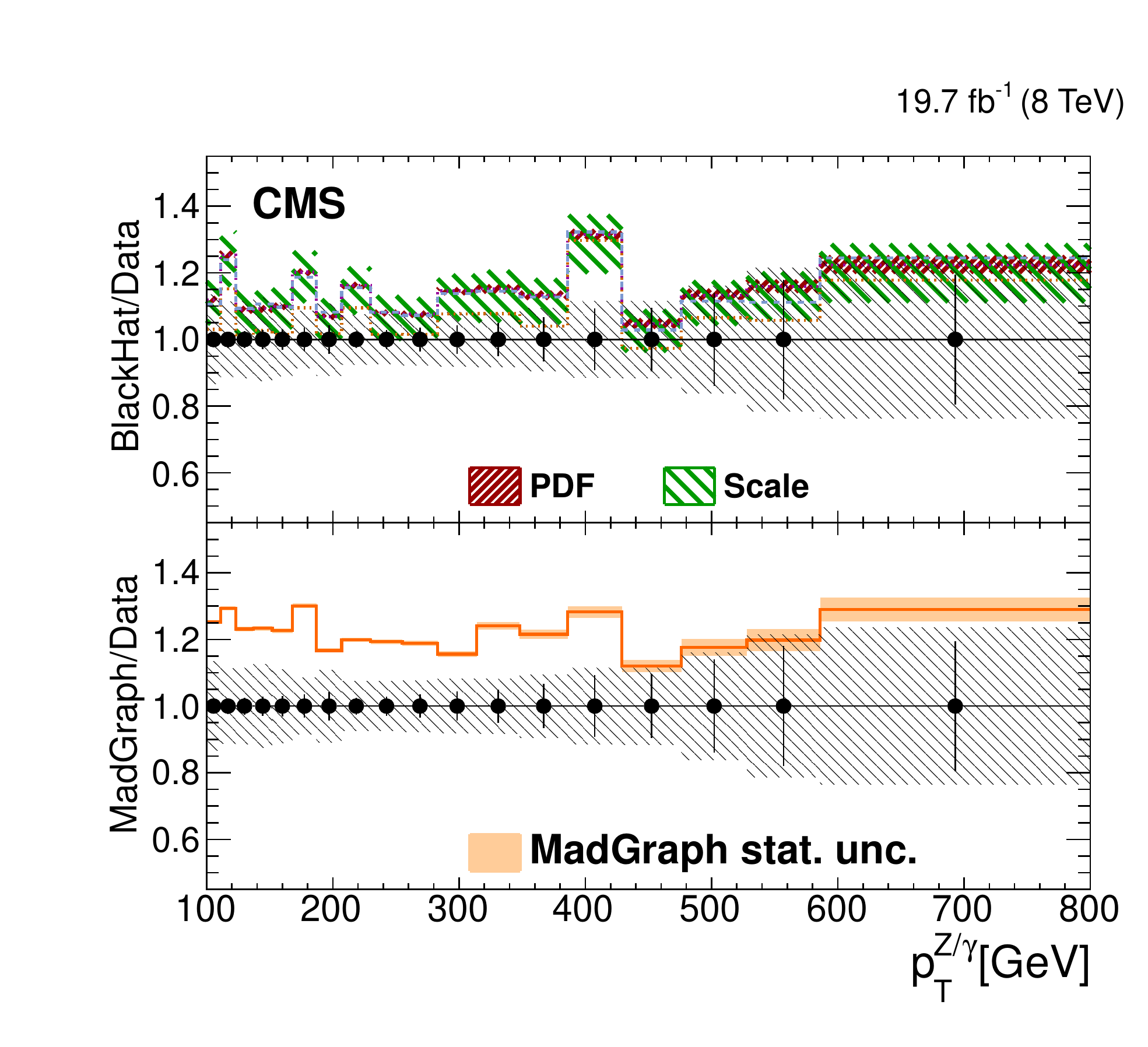}
  \includegraphics[width=0.49\textwidth]{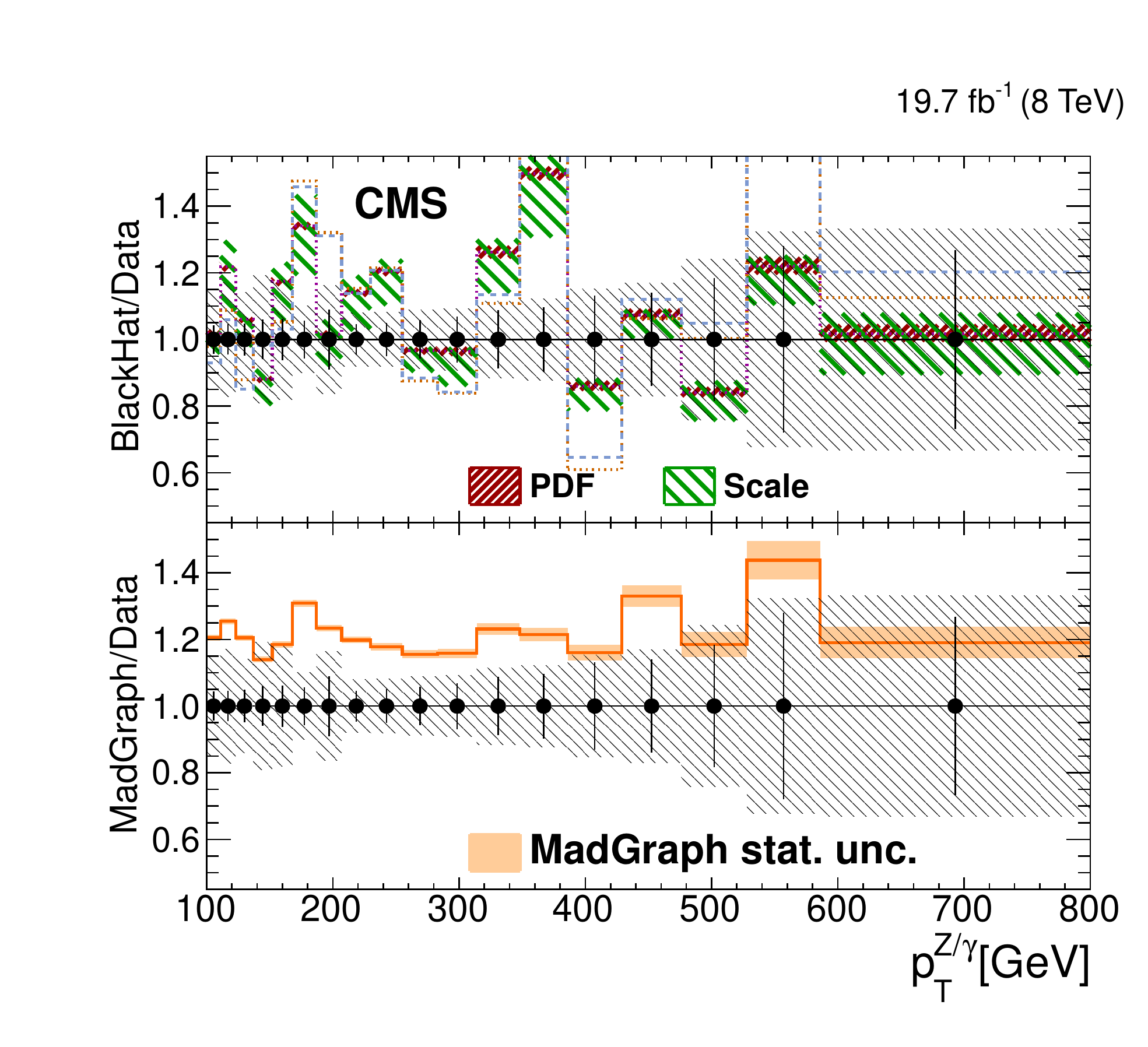}
  \caption{Differential cross section ratio of $\cPZ\to\left(\Pep\Pem+\PGmp\PGmm\right)$ over  $\gamma$  as a function
of the total transverse-momentum cross section and for central bosons ($\abs{y^{V}}<1.4$) at different kinematic selections in detector-corrected data. Top left: 2-jet ($\njets \ge 2$); top right: 3-jet ($\njets\ge3$). The black error bars reflect the statistical uncertainty in the ratio, the hatched (gray) band represents the total uncertainty in the measurement. The shaded band around the \MADGRAPHPYTHIASIX simulation to data ratio represents the statistical uncertainty in the MC estimation. The bottom plots give the ratio of the various theoretical estimations to the data in the $\njets\geq2$ case (bottom left) and $\njets\geq3$ case (bottom right).}
   \label{fig:Zpt_over_gamma_pt_1}
\end{figure}

The resulting ratio distributions are shown in Figs.~\ref{fig:Zpt_over_gamma_pt} and~\ref{fig:Zpt_over_gamma_pt_1} for all selections.
The $\zjets$ selection with the requirement $\HT>300\GeV$ enhances the presence of
events with large hadronic activity.

In all phase space regions, we observe a ratio which saturates around $\pt\simeq 300\--350\GeV$. This agrees with the LO estimations stating that the main distinction between the two processes is the mass difference, with the second difference being the different couplings.

In the inclusive $\njets\geq1$ selection (Fig.~\ref{fig:Zpt_over_gamma_pt}), the plateau value is
\begin{equation}
  R_\text{dilep}=\frac{\sigma_{\cPZ\to \ell^{+}\ell^{-}} (\pt^{\cPZ}>314\GeV)}{\sigma_{\gamma}(\pt^{\gamma}>314\GeV)}=0.0322\pm0.0008\stat\pm0.0020\syst.
\end{equation}
Here $R_\text{dilep}$ is the plateau value of the ratio of the dilepton $\cPZ$ cross section and the \gjets~cross section for the last seven bins ($\pt^V>314\GeV$). This translates into the ratio of the total cross sections of $R_\text{tot}=0.957\pm0.066$ when divided by the average leptonic branching fraction of $(3.3658\pm0.0023)\%$~\cite{PDGreport}.

The estimation from \MADGRAPHPYTHIASIX is overlaid in Figs.~\ref{fig:Zpt_over_gamma_pt} and ~\ref{fig:Zpt_over_gamma_pt_1}, where
the LO estimation is used to compare $\zjets$ and $\gjets$ differential cross sections at the same order of perturbative expansion.
Although \MADGRAPHPYTHIASIX does not reproduce the high-\pt tail for either $\zjets$ or $\gjets$, the shapes of the curves are similar for both processes and their ratio is flat. Using LO cross sections, \MADGRAPHPYTHIASIX predicts a ratio with a value of $R_{\mathrm{MG}}=0.0391$, which is higher than that observed in data by a factor of $1.21\pm0.08$ (stat+syst). No clear trend away from a flat ratio is observed.
Higher-order effects beyond LO, which could lead to a rise or fall in the plateau region, are smaller than the experimental uncertainties.

The \BLACKHAT estimation is also overlaid in Figs.~\ref{fig:Zpt_over_gamma_pt}--\ref{fig:Zpt_over_gamma_pt_1} and reproduces the 1-jet and 2-jet ratio to within 10\% across the entire range. It reproduces the $\HT\geq300\GeV$ case accurately in the low-\pt regime and results in an approximately 20\% overestimation in the high-\pt range. In the region where $\pt^V<300\GeV$, the scale uncertainty grows to roughly 30\%. This corresponds to the region where \BLACKHAT fails to reproduce the $\ptZ$ and $\ptG$ spectra separately. Inclusive fixed-order calculations are not designed to model this selection of high jet activity with a comparatively low boson \pt. In the 3-jet case, \BLACKHAT overestimates the ratio by approximately 25\%, but agrees with data starting around the plateau region of approximately 300\GeV.

We calculate the scale and PDF uncertainty bands for \BLACKHAT using the scale and PDF uncertainty envelopes from the $\ptZ$ and $\ptG$ spectra. If we correlate the different renormalization and factorization scales ($\mu_R$ and $\mu_F$), the envelope decreases to approximately 2\%, whereas if we take the scales as completely anticorrelated, we see a band of approximately 10\% in the bulk. However, we know that the former underestimates the theoretical uncertainty due to renormalization and factorization scales, and the latter overestimates it. The estimation of this uncertainty has been discussed in the literature, and has been examined by comparing different theoretical computational estimations (\cite{BlackHat} and~\cite{Bern:2011pa}). Both of the previously mentioned methods misrepresent the actual uncertainty due to the renormalization and factorization scales. We therefore choose the larger relative scale uncertainty band from each process as an estimate of the uncertainty on the final ratio.
 Using the NLO cross sections, \BLACKHAT predicts the $R_\text{dilep}$ ratio with a value of $R_{\mathrm{BH}}=0.03794$, which is higher than that observed in data by a factor of $1.18\pm0.14\, (\text{stat}+\text{syst})$.

\section{Summary}
\label{sec:conclusions}
Differential cross sections have been measured for $\zjets$ (with $\Z\to \ell^{+}\ell^{-}$) and isolated $\gjets$ as a function of the boson transverse momentum, using data collected by CMS at $\sqrt{s}=8\TeV$ corresponding to an integrated luminosity of 19.7\fbinv. The estimations from the MC multiparton LO+PS generators \MADGRAPHPYTHIASIX and \SHERPA have been compared to the data. We find that the $\pt$ spectra for $\zjets$ and $\gjets$ are not well reproduced by these MC models. We observe a monotonic increase of the MC simulation/data ratio with increasing vector boson \pt. Using the NLO generator \BLACKHAT simulation, we find a smaller discrepancy in shape between data and simulation, indicating that it is likely related to missing higher-order effects.

We have also studied the distribution of the ratios of $\pt^{\cPZ}$ and hadronic quantities ($\HT$ and $\pt^{\mathrm{j}1}$) in $\zjets$. We find that these agree with the LO+PS estimation over the whole range when an NNLO $K$-factor is applied. The NLO \BLACKHAT estimation is accurate in a subrange where the NLO estimation is expected to perform well.

In addition, we presented a measurement of the ratio of the $\zjets$ to $\gjets$ cross sections in four phase space regions: $\njets \geq 1,$ 2, 3, and $\HT>300\GeV$, $\njets \geq 1$.
\MADGRAPHPYTHIASIX (LO+PS) overestimates the data by a factor $1.21\pm0.08$ (stat+syst), whereas \BLACKHAT (NLO) overestimates the data by a factor $1.18\pm0.14$ (stat+syst) in the plateau region i.e., for $\pt^{V}$ above approximately 300\gev. As a function of the vector boson transverse momentum, these factors are at similar values of around 1.2 for all the considered phase space selections. Thus, we find that simulations reproduce the shape of the ratio of $\ptZ$ to $\ptG$ distributions better than the individual $\ptZ$ or $\ptG$ distributions in all selections considered. These four selections mimic phase space regions of interest for searches of physics beyond the standard model. We emphasize that the agreement is similar for different jet multiplicities and $\HT$ ranges because \zjets{} and \gjets{} events have been generated with the same level of accuracy for up to four partons in the final-state ME. In the comparison, we considered both processes at either LO or at NLO. It is clear from the differences observed between the NLO and LO+PS estimations in each process, the conclusions may not be true if the samples are generated with different orders of accuracies of the matrix element calculation.

Our results show that properties of the $\Znunu$ process can be predicted using the measured \gjets~final state and the simulated ratio between $\Znunu+\text{jets}$ and $\gjets$. However, this simulated ratio must be corrected with the measured ratio of leptonic $\zjets$ and $\gjets$.

\begin{acknowledgments}
We congratulate our colleagues in the CERN accelerator departments for the excellent performance of the LHC and thank the technical and administrative staffs at CERN and at other CMS institutes for their contributions to the success of the CMS effort. In addition, we gratefully acknowledge the computing centers and personnel of the Worldwide LHC Computing Grid for delivering so effectively the computing infrastructure essential to our analyses. Finally, we acknowledge the enduring support for the construction and operation of the LHC and the CMS detector provided by the following funding agencies: BMWFW and FWF (Austria); FNRS and FWO (Belgium); CNPq, CAPES, FAPERJ, and FAPESP (Brazil); MES (Bulgaria); CERN; CAS, MoST, and NSFC (China); COLCIENCIAS (Colombia); MSES and CSF (Croatia); RPF (Cyprus); MoER, ERC IUT and ERDF (Estonia); Academy of Finland, MEC, and HIP (Finland); CEA and CNRS/IN2P3 (France); BMBF, DFG, and HGF (Germany); GSRT (Greece); OTKA and NIH (Hungary); DAE and DST (India); IPM (Iran); SFI (Ireland); INFN (Italy); MSIP and NRF (Republic of Korea); LAS (Lithuania); MOE and UM (Malaysia); CINVESTAV, CONACYT, SEP, and UASLP-FAI (Mexico); MBIE (New Zealand); PAEC (Pakistan); MSHE and NSC (Poland); FCT (Portugal); JINR (Dubna); MON, RosAtom, RAS and RFBR (Russia); MESTD (Serbia); SEIDI and CPAN (Spain); Swiss Funding Agencies (Switzerland); MST (Taipei); ThEPCenter, IPST, STAR and NSTDA (Thailand); TUBITAK and TAEK (Turkey); NASU and SFFR (Ukraine); STFC (United Kingdom); DOE and NSF (USA).

Individuals have received support from the Marie-Curie program and the European Research Council and EPLANET (European Union); the Leventis Foundation; the A. P. Sloan Foundation; the Alexander von Humboldt Foundation; the Belgian Federal Science Policy Office; the Fonds pour la Formation \`a la Recherche dans l'Industrie et dans l'Agriculture (FRIA-Belgium); the Agentschap voor Innovatie door Wetenschap en Technologie (IWT-Belgium); the Ministry of Education, Youth and Sports (MEYS) of the Czech Republic; the Council of Science and Industrial Research, India; the HOMING PLUS program of the Foundation for Polish Science, cofinanced from European Union, Regional Development Fund; the Compagnia di San Paolo (Torino); the Consorzio per la Fisica (Trieste); MIUR project 20108T4XTM (Italy); the Thalis and Aristeia programs cofinanced by EU-ESF and the Greek NSRF; and the National Priorities Research Program by Qatar National Research Fund.
\end{acknowledgments}

\addcontentsline{toc}{section}{\refname}
\bibliography{auto_generated}

\cleardoublepage \appendix\section{The CMS Collaboration \label{app:collab}}\begin{sloppypar}\hyphenpenalty=5000\widowpenalty=500\clubpenalty=5000\textbf{Yerevan Physics Institute,  Yerevan,  Armenia}\\*[0pt]
V.~Khachatryan, A.M.~Sirunyan, A.~Tumasyan
\vskip\cmsinstskip
\textbf{Institut f\"{u}r Hochenergiephysik der OeAW,  Wien,  Austria}\\*[0pt]
W.~Adam, E.~Asilar, T.~Bergauer, J.~Brandstetter, E.~Brondolin, M.~Dragicevic, J.~Er\"{o}, M.~Flechl, M.~Friedl, R.~Fr\"{u}hwirth\cmsAuthorMark{1}, V.M.~Ghete, C.~Hartl, N.~H\"{o}rmann, J.~Hrubec, M.~Jeitler\cmsAuthorMark{1}, V.~Kn\"{u}nz, A.~K\"{o}nig, M.~Krammer\cmsAuthorMark{1}, I.~Kr\"{a}tschmer, D.~Liko, I.~Mikulec, D.~Rabady\cmsAuthorMark{2}, B.~Rahbaran, H.~Rohringer, J.~Schieck\cmsAuthorMark{1}, R.~Sch\"{o}fbeck, J.~Strauss, W.~Treberer-Treberspurg, W.~Waltenberger, C.-E.~Wulz\cmsAuthorMark{1}
% ...
\vskip\cmsinstskip
% ...
1:~~Also at Vienna University of Technology, Vienna, Austria\\
2:~~Also at CERN, European Organization for Nuclear Research, Geneva, Switzerland\\
% ...

\end{sloppypar}
\end{document}